\documentclass{article}

\usepackage{PRIMEarxiv}

\usepackage[utf8]{inputenc} 
\usepackage[T1]{fontenc}    
\usepackage{hyperref}       
\usepackage{url}            
\usepackage{booktabs}       
\usepackage{amsfonts}       
\usepackage{nicefrac}       
\usepackage{microtype}      
\usepackage{lipsum}
\usepackage{fancyhdr}       
\usepackage{graphicx}       
\graphicspath{{media/}}     
\usepackage[square,numbers]{natbib}
\usepackage{setspace}
\usepackage{indentfirst}
\usepackage{float}
\usepackage{caption}
\usepackage[belowskip=-10pt,aboveskip=5pt]{caption}
\usepackage{subcaption}
\usepackage{relsize}
\usepackage{mathrsfs,amsmath}
\usepackage{mathtools}
\usepackage{algorithm}
\usepackage{algpseudocode} 
\usepackage{color,soul}
\usepackage{tabularx}
\usepackage{diagbox}

\hypersetup{
	colorlinks=true,
	linkcolor=blue,
	citecolor=blue,
	urlcolor=blue,
	pdftitle={Process Optimization under Uncertainty for Improving the Bond Quality of Polymer Filaments in Fused Filament Fabrication},
	pdfkeywords={fused filament fabrication, polymer additive manufacturing, process optimization, uncertainty quantification, bond quality},  
	pdfauthor={Berkcan Kapusuzoglu, Matthew Sato, Sankaran Mahadevan, Paul Witherell},
}

\title{Process Optimization under Uncertainty for Improving the Bond Quality of Polymer Filaments in Fused Filament Fabrication
\thanks{\textit{\underline{Citation}}: 
\textbf{Kapusuzoglu, B., Sato, M., Mahadevan, S., \& Witherell, P. Process optimization under uncertainty for improving the bond quality of polymer filaments in fused filament fabrication. \textit{Journal of Manufacturing Science and Engineering} 143(2), 021004 (2021). DOI: \href{https://doi.org/10.1115/1.4048380}{10.1115/1.4048380}}
} 
}

\author{
  \textbf{Berkcan Kapusuzoglu}\thanks{Corresponding author: berkcan.kapusuzoglu@vanderbilt.edu}$^{\ ,1}$, \textbf{Matthew Sato}$^1$, \textbf{Sankaran Mahadevan}$^1$, \textbf{Paul Witherell}$^2$ \\
  \vspace{0.2cm} \\
  $^1$Department of Civil and Environmental Engineering, Vanderbilt University, Nashville, TN 37235, USA \\
  $^2$Systems Integration Division, Engineering Laboratory, \\ National Institute of Standards and Technology (NIST), Gaithersburg, MD 20899, USA
}

\begin{document}
\maketitle

\begin{abstract}
This paper develops a computational framework to optimize the process parameters such that the bond quality between extruded polymer filaments is maximized in fused filament fabrication (FFF). A transient heat transfer analysis providing an estimate of the temperature profile of the filaments is coupled with a sintering neck growth model to assess the bond quality that occurs at the interfaces between adjacent filaments. Predicting the variability in the FFF process is essential for achieving proactive quality control of the manufactured part; however, the models used to predict the variability are affected by assumptions and approximations. This paper systematically quantifies the uncertainty in the bond quality model prediction due to various sources of uncertainty, both aleatory and epistemic, and includes the uncertainty and the model discrepancy in the process parameter optimization. Variance-based sensitivity analysis based on Sobol' indices is used to quantify the relative contributions of the different uncertainty sources to the uncertainty in the bond quality. A Gaussian process (GP) surrogate model is constructed to compute and include the model discrepancy within the optimization. Physical experiments are conducted for calibration and validation of the physics model, and also for validation of the optimum solution. The results show that the proposed formulation for process parameter optimization under uncertainty results in high bond quality between adjoining filaments of the FFF product.
\end{abstract}


\section{Introduction}
Fused filament fabrication (FFF), an extrusion-based deposition technique, is a widely used additive manufacturing (AM) process. Among other rapid prototyping technologies, FFF is popular due to its low cost, easy operation, and suitability for complex geometries. FFF is the process of joining materials, usually layer upon layer, by extruding a molten filament through a heated nozzle at a controlled rate and a gantry, which moves in the horizontal plane in a predefined pattern onto a build plate or onto other filaments that move in the vertical direction. The build plate supporting the extruded polymer filaments is typically set at a controlled temperature.

\begin{figure}[!htb]
\centering\includegraphics[width=0.45\linewidth]{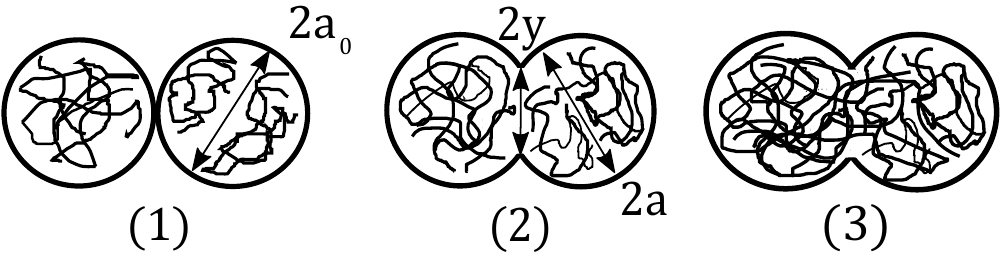}
\caption{The bond formation process between two filaments:~(1) initial surface contact;~(2) wetting or neck growth;~(3) molecular diffusion and randomization across the cross-section of two FFF extruded filaments} 
	\label{fig:BondProcess}
\end{figure}

The material cools down, solidifies and bonds with the surrounding filaments as it is deposited. The bond formation process in FFF, as illustrated in Fig.~\ref{fig:BondProcess} is mainly affected by the thermal energy of the extruded material. The bond quality is mainly driven by wetting, also known as the neck growth phenomenon, molecular diffusion, and randomization when the interface temperature is above the critical sintering temperature (Fig.~\ref{fig:BondProcess}). The neck growth between adjacent filaments within a layer may be termed as intra-layer bonding, and the similar neck growth evolution that occurs between two successive layers may be called as inter-layer bonding. As an additional measure, inter-layer adhesion of the printed parts can be used to analyze the quality of inter-layer bonding. There are some studies using fracture-mechanics-based methods, which characterize the fracture resistance and the inter-layer adhesion of FFF 3D printed materials~\cite{aliheidari2017fracture}.

The bond quality between adjacent filaments and layers strongly affects the mechanical properties of FFF-produced parts. The correlation between better neck radius and overall part strength has been studied by Sun et al.~\cite{sun2008effect} where it is shown that that better bonding between filaments results in greater mechanical strength. The temperature history at the interfaces between filaments has a direct impact on the bond quality and plays an important role in predicting FFF specimen strength since it affects the neck growth, molecular diffusion and thermal stresses. Thus, it is important to determine the evolution of temperature in the interfaces between filaments in order to predict the bond formation, and therefore the mechanical properties of the manufactured part. Yardimci et al.~\cite{yardimci1996part} and Yardimci and G{\"u}{\c{c}}eri~\cite{atif1996conceptual} presented numerical heat transfer models for fused filament fabrication of ceramics. They modeled the cooling process of a filament due to convection with the environment and compared different build patterns without taking contacts with adjoining filaments into account. Thomas and Rodr{\'\i}guez~\cite{thomas2000modeling} modeled the fracture strength of the FFF part by developing a transient 2D analysis using a finite element method. They assumed the cross section of the filaments to be rectangular and neglected contact resistances. Li et al.~\cite{li2002investigation} proposed a 1D analytical transient heat transfer model coupled with a lumped capacity method, considering elliptical filaments with a semi-infinite filament length, which means that temperature is uniform across the cross-section, and varies along the length of the filaments. Sun et al.~\cite{sun2008effect} compared the heat transfer models developed by Li et al.~\cite{li2002investigation} and Thomas and Rodr{\'\i}guez~\cite{thomas2000modeling} to the temperature of the bottom-most filament using thermocouples. They found that the model proposed by Li et al.~\cite{li2002investigation} underestimated conduction, and the model developed by Thomas and Rodr{\'\i}guez~\cite{thomas2000modeling} underestimated convection. They assessed the bond quality both experimentally and using a sintering neck growth model, and concluded that the sintering phenomenon had a significant effect on bond strength development only for a short time since the temperature profile of the filaments remained below the critical sintering temperature after a very short time. Costa et al.~\cite{costa2015thermal} analyzed the mechanisms that had the largest effect on the cooling process of filaments. They found convection with the environment, and conduction with adjacent filaments were the most significant mechanisms. Costa et al.~\cite{costa2017estimation} proposed a transient heat transfer model that considers the physical contacts between a filament and its adjacent filaments or the build plate. Current models in the literature do not include the humidity in the environment, the effect of filament age and percentage of humidity in the filament.

The sintering process, describing the neck growth between filaments (stage 2 in Fig.~\ref{fig:BondProcess}), is defined as the coalescence of particles under the action of surface tension~\cite{pokluda1997modification}. A Newtonian sintering model for polymers was initially developed by Frenkel et al.~\cite{frenkel1945viscous} to predict the rate of polymer sintering. Pokluda et al.~\cite{pokluda1997modification} developed a closed-form equation to predict the neck growth between two spherical particles based on the work balance of viscous dissipation and surface tension. Bellehumeur et al.~\cite{bellehumeur2004modeling} applied the model proposed by Pokluda et al.~\cite{pokluda1997modification} to FFF for predicting the sintering between adjacent filaments as a nonlinear function of time, temperature-dependent material properties, and an initial spherical particle radius. Gurrala et al.~\cite{gurrala2014part} extended the study done by Bellehumeur et al.~\cite{bellehumeur2004modeling} by considering cylindrical filament geometry rather spherical particles.

The models used to predict the bond quality are affected by various sources of uncertainty and error. This affects process optimization and process control decisions. Therefore, it is critical to identify and quantify the effects of the uncertainty and error sources in order to improve the overall quality of FFF products. Many previous studies on process optimization in AM have employed a design of experiments-based trial-and-error approach to determine the effect of different process parameter settings on the quantity of interest such as part strength, residual stress, surface roughness etc.~\cite{sun2008effect,kobryn2001laser,rodriguez2000characterization,pierson2018process}. The experiment-based approach is problem-specific, and cannot be used as a general-purpose process design method for different materials and part geometries. Therefore, there is a need to use a physics model-based approach that optimizes the process parameters for better overall bond quality without running multiple economically expensive physical experiments, and also includes the uncertainty in the neck growth prediction.

The focus of this work is to develop a framework for optimizing process parameters under uncertainty that maximizes the filament bond quality of FFF printed parts. The overall bond length is the decision criterion we use for optimization; greater overall bond lengths signify lesser void area (thus better bonding) in an average sense. We only consider stage 1 and 2 shown in Fig.~\ref{fig:BondProcess} to predict the overall bond length using the sintering neck growth model shown in Section~\ref{sec:bondlength}. Stage 3 is not considered in the bond formation modeling. Other alternative criteria could also be used within this framework, such as reducing the void area. Note that the focus of this paper is not to improve the process, but to make optimum use of the existing process. In this context, the proposed methodology leverages the transient heat transfer model and sintering neck growth model for cylindrical filaments proposed by Costa et al.~\cite{costa2015thermal} and Gurrala et al.~\cite{gurrala2014part}, respectively. It quantifies the uncertainty in the coupled multi-physics model prediction, and determines the optimal process parameters through model-based optimization under uncertainty to improve the bond quality at each layer to overcome poor intra-layer and inter-layer bonding. Monte Carlo simulation (MCS) is used to propagate uncertain quantities through the coupled physics-based models by generating random samples from the design space. Optimization under uncertainty requires the evaluation of the uncertainty in the output for each iteration of the optimization algorithm. In other words, the MCS is nested within the optimization step, making the optimization under uncertainty very expensive. Therefore, a surrogate-based optimization framework~\cite{gutmann2001radial} is employed, where an inexpensive surrogate of the original physics model is used to find the optimum process parameters. 

In summary, the contributions of this paper are as follows:
\begin{enumerate}
    \item Development of a Bayesian methodology to quantify the uncertainty in the neck growth prediction.
    \item Construction of a GP surrogate model for efficient computation of model error, in order to incorporate model error in process parameter optimization.
    \item Evaluation of the relative contributions of various uncertainty sources to the uncertainty in the model output.
    \item Development of a computational framework for process parameter optimization under uncertainty, in order to maximize the bond quality between extruded polymer filaments in FFF.
\end{enumerate}

The outline of the rest of this paper is as follows. Section~\ref{Sec:Background} provides background information on physics-based models in FFF. Section~\ref{Sec:Methods} presents the basic framework of Bayesian model calibration, variance-based sensitivity analysis, surrogate modeling, and develops the proposed optimization under uncertainty methodology that optimizes the process parameters of FFF for maximizing the overall bond quality of the part. The proposed methodology is illustrated for a numerical example and is validated with physical experiments in Section~\ref{Sec:Results}. Concluding remarks are provided in Section~\ref{Sec:Conclusion}.

\section{Background}\label{Sec:Background}
\subsection{Heat transfer analysis}
\begin{figure}[!htb]
\centering\includegraphics[width=0.4\linewidth]{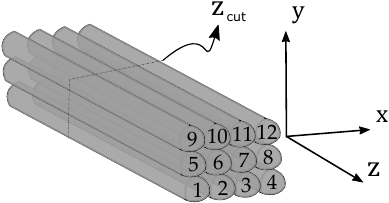}
\caption{Example of a filament deposition sequence} 
	\label{fig:ExampleDepoSeq}
\end{figure}

In FFF, all the filaments are subjected to the same heat transfer mechanism, but with different boundary conditions depending on the part geometry and deposition sequence (Fig.~\ref{fig:ExampleDepoSeq}). The heat transfer model needs to be run for each interface since the neck growth predictions between each filament require the temperature evolution as an input. The time-dependent temperature profiles of filaments can be determined by performing a one-dimensional transient heat transfer analysis developed by Costa et al.~\cite{costa2017estimation}. The reason why a low-fidelity model has been used instead of a 2D or 3D high-fidelity model is that the low-fidelity model is much cheaper to run. Moreover, the low-fidelity model does well at higher temperatures, where most of the neck growth occurs.

\begin{figure*}[!htb]
\centering
\begin{subfigure}[!htb]{0.45\textwidth} 
\centering{
  \includegraphics[width=.8\columnwidth,keepaspectratio]{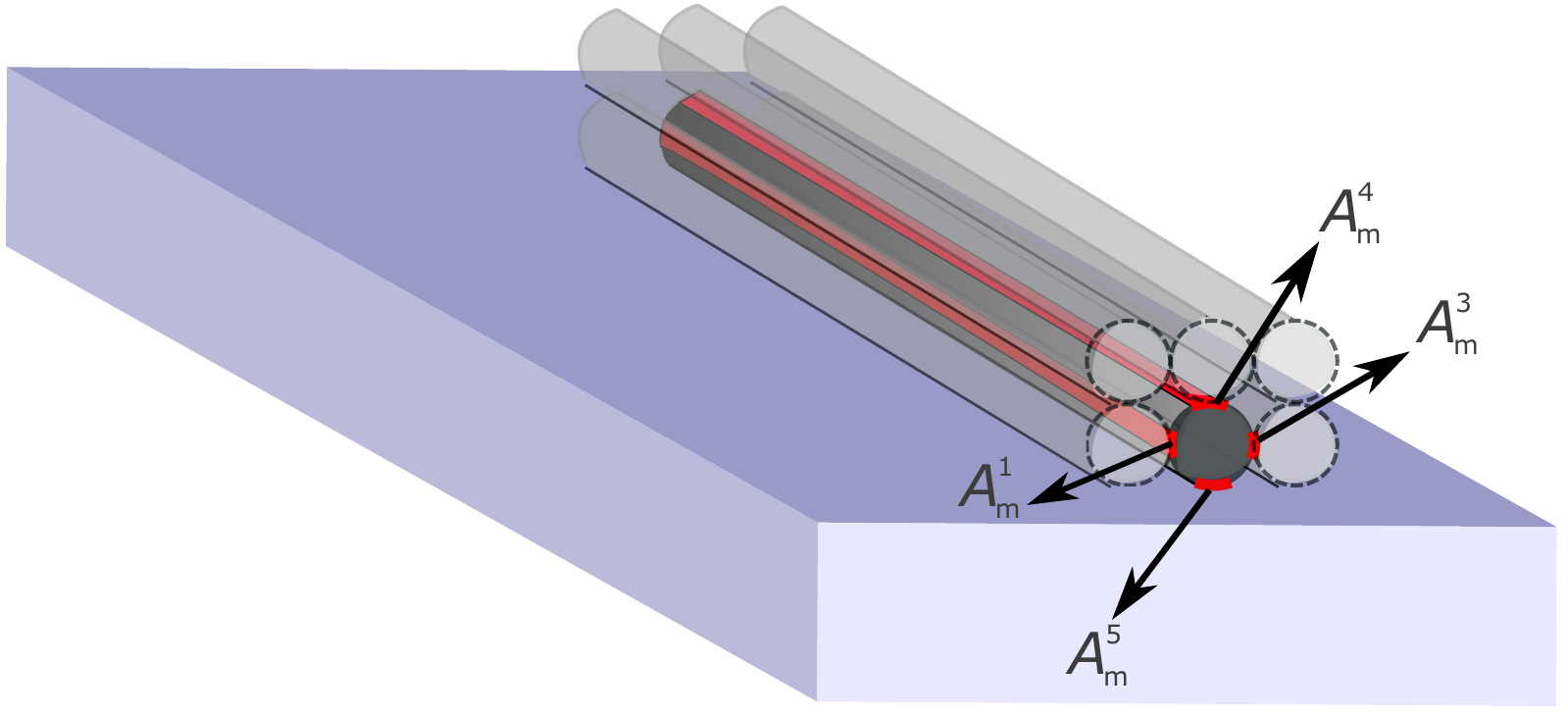}
}
\subcaption{}
\end{subfigure}%
\begin{subfigure}[!htb]{0.45\textwidth} 
\centering{
\includegraphics[width=.8\columnwidth,keepaspectratio]{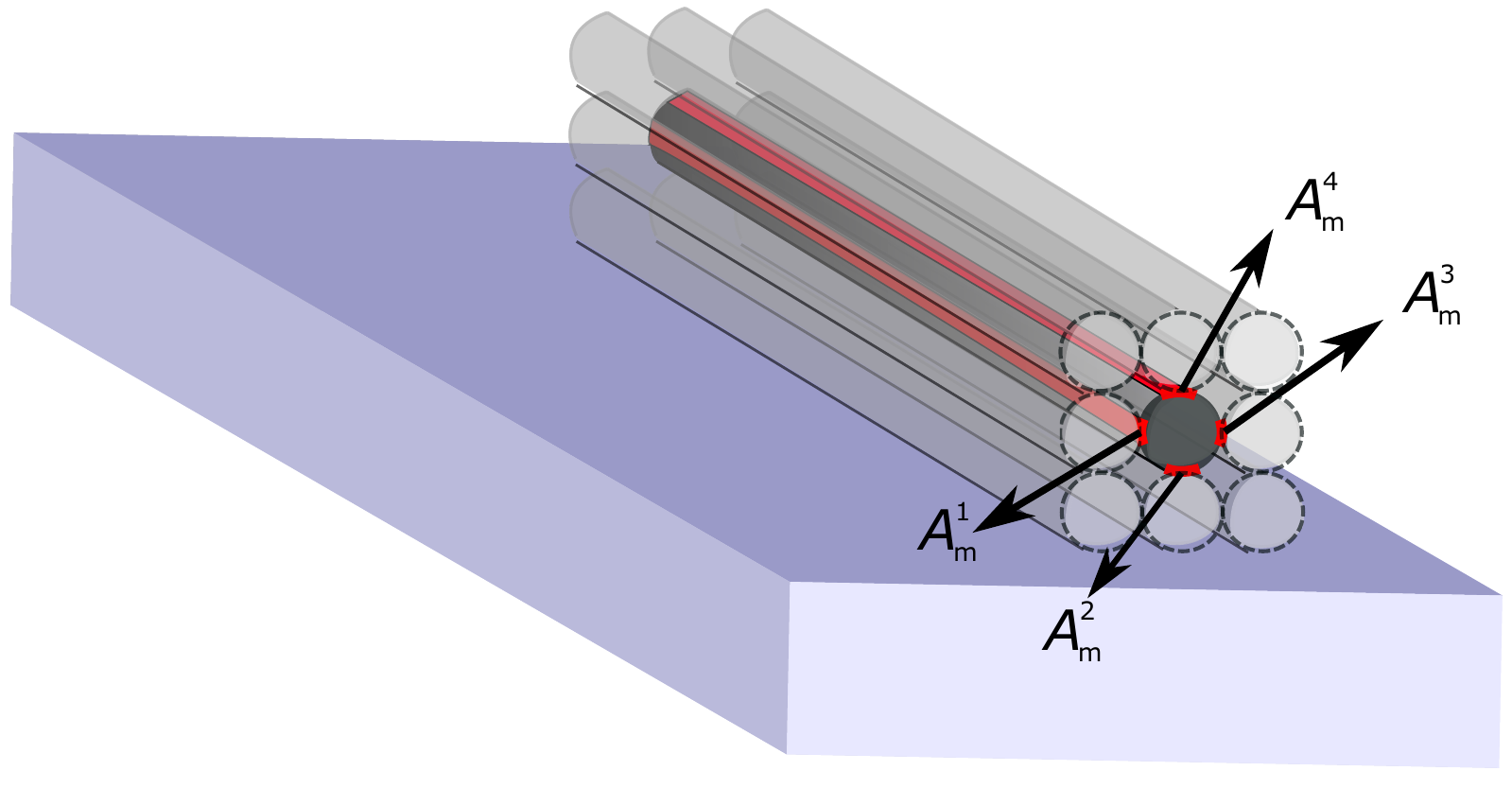}
\subcaption{}
}\end{subfigure}%
\caption{Possible contact areas of a filament in the:~(a) first layer and~(b) remaining layers}
\label{fig:ContactAreas}
\end{figure*}

The deposition of filaments is modeled gradually by joining elementary lengths that are associated with a given deposition time. The mathematical energy balance for an elementary length $dx$ can be written as:
\begin{align}
\small
    -k A\frac{\partial T_m(x,t)}{\partial x} - {}& h_{conv} A_m^{conv} \left(T_m(x,t)-T_{S}\right) -  \sum_{i=1}^n h_i A_m^i \left(T_m(x,t)-T_m^i \right) = \alpha A \frac{\partial T_m(x,t)}{\partial t}dx - \notag\\
    {}& A \left[ k \frac{\partial T_m(x,t)}{\partial x} + \dfrac{\partial \left(k\dfrac{\partial T_m(x,t)}{\partial x} \right)}{\partial x}dx \right] 
    \label{eq:EnergyBalance}
\end{align}
where $k$, $h_{conv}$, $h_i$ and $\alpha=\rho C$ are the material properties of the polymer and assumed to be temperature-independent. The parameter $k$ is the thermal conductivity, $h_{conv}$ is the convective heat transfer coefficient, $h_i$ represents the contact heat transfer coefficient, $\rho$ and $C$ are the density and specific heat capacity of the material, $T_m$ is the temperature at cross-section $z_{\rm cut}$ of the $m$-th filament ($m\in \{1,...,N\}$, where $N$ is the total number of deposited filaments) at deposition time instant $t$, $T_m^i$ represents the temperature of the adjacent filament or build plate at contact $i$ ($i\in \{1,...,n\}$, where $n$ is the total number of contact surfaces of a filament including the contact with the build plate) of the $m$-th filament or the build plate, $T_{S}$ is the surrounding environment temperature, and $A$ represents the cross-section area of a filament. $A_m^{conv}$ is the area of the $m$-th filament that is in contact with the environment, $A_m^i$ is the area of contact $i$ for the $m$-th filament as shown in Fig.~\ref{fig:ContactAreas}, and they are defined as:
\begin{align}
    A_m^{conv}  =\ & P \left( 1 - \sum_{i=1}^n \omega_m^i \lambda_i \right) dx, \\
    & A_m^i = P \omega_m^i \lambda_i\ dx,
\end{align}
where $P$ is the filament perimeter, $\lambda_i$ is the fraction of $P$ that is in contact with another filament or with the build plate, and $\omega_m^i$ is a variable, which equals unity if the $m$-th filament has the $i$-th contact, and zero otherwise. 

Axial and radial heat conduction can be neglected due to the low thermal conductivity of polymers and small filament radius~\cite{costa2008towards}. Humidity in the environment, the effect of filament age and percentage of humidity in the filament were not considered in the heat transfer analysis. Thus, after these assumptions the energy equation becomes:
\begin{align}
\small
    \frac{\partial T_m(x,t)}{\partial t} = - \frac{P}{\alpha A} {}& \Bigg[ h_{conv} \left( 1-\sum_{i=1}^n \omega_m^i \lambda_i \right) \big( T_m(x,t)-T_{BP} \big) + \sum_{i=1}^n h_i \omega_m^i\lambda_i \left(T_m(x,t)- T_m^i\right) \Bigg].
    \label{eq:EnergyBalanceFinal}
\end{align}
The analytical solution of Eq.~\eqref{eq:EnergyBalanceFinal} can be obtained using the characteristic polynomial method \cite{palais2009differential}:
\begin{equation}
\small
    T_m(x,t) = \phi_1\ \text{exp} \left[\dfrac{P \chi \left(\omega_m^1,...,\omega_m^n\right)}{\alpha A} \big(t-t_m(x) \big)\right] + \psi\left(\omega_m^1,...,\omega_m^n\right),
\end{equation}
where $\phi_1=T_m(t_m(x))-\psi\left(\omega_m^1,...,\omega_m^n\right)$, $T_m(t_m(x))$ is the temperature of the $m$-th filament at instant $t_m(x)$ at which an elementary length $x$ of the $m$-th filament is deposited and starts to cool down or contact with an adjacent filament or the build plate. The functions that are influenced by the contacts $\chi$ and $\psi$ are defined as:
\begin{equation}
\small
    \chi\left(\omega_m^1,...,\omega_m^n\right) = h_{conv}\left(1-\sum_{i=1}^n \omega_m^i\lambda_i\right) + \sum_{i=1}^n \omega_m^i h_i\lambda_i,
\end{equation}
\begin{equation}
\small
    \psi\left(\omega_m^1,...,\omega_m^n\right) = \dfrac{h_{conv}\left(1-\sum_{i=1}^n \omega_m^i\lambda_i\right) T_E + \sum_{i=1}^n \omega_m^i h_i\lambda_i T_m^i}{\chi\left(\omega_m^1,...,\omega_m^n\right)}.
\end{equation}
\subsection{Sintering neck growth model between filaments}\label{sec:bondlength}
In this paper, we consider a sintering neck growth model to assess the bond quality that occurs at the interfaces between adjacent filaments. Note that bond length measurements are at the macro-scale; therefore our computations have also been at this scale. The bonding is quantified by measuring the length of the perimeter that is shared by adjacent filaments. A Newtonian sintering model was initially developed by Frenkel~\cite{frenkel1945viscous}. Pokluda et al.~\cite{pokluda1997modification} developed a closed-form equation to predict the bond length (half of which is neck radius) $y$ as shown in Fig.~\ref{fig:BondProcess}.

Bellehumeuer et al.~\cite{bellehumeur2004modeling} applied the model proposed by Pokluda et al.~\cite{pokluda1997modification} to FFF to predict the neck growth between adjacent filaments as a nonlinear function of time $t$, temperature-dependent surface tension $\Gamma$, viscosity of the ABS material $\eta$, and an initial radius $a_0$ of the extruded filament before the sintering process. In this work, the Newtonian sintering model illustrated in Eq.~\eqref{eq:SinteringModel} is used to predict the intra-layer bond lengths of the FFF product.
\begin{equation}
    \frac{d\theta}{dt} = \frac{\Gamma}{a_0 \eta} \frac{2^{-5/3} \mathrm{cos}\ \theta\ \rm  sin\ \theta (2-\mathrm{cos}\ \theta)^{1/3}}{(1-\mathrm{cos}\ \theta)(1+\mathrm{cos}\ \theta)^{1/3}}
    \label{eq:SinteringModel}
\end{equation}
where $\theta=\mathrm{sin}^{-1}y/a$ is the bond angle.

The bonding model (Eq.~\eqref{eq:SinteringModel}) assumes that the filaments are spheres. In order to predict the bond lengths between two cylindrical filaments, a model developed by Gurrala et al.~\cite{gurrala2014part} given in Eq.~\eqref{eq:SinteringModelCylind} is used:
\begin{equation}
\small
    \frac{d\theta}{dt} = \frac{\Gamma}{3\sqrt{\pi}a_0 \eta} \frac{[(\pi-\theta) \mathrm{cos}\ \theta + \mathrm{sin}\ \theta][\pi-\theta + \mathrm{sin}\ \theta\ \mathrm{cos}\ \theta]^{1/2}}{(\pi-\theta)^2\ \mathrm{sin}^2\ \theta}
    \label{eq:SinteringModelCylind}
\end{equation}

\section{Proposed Methodology}\label{Sec:Methods}
The proposed methodology for process parameter optimization under uncertainty consists of the following steps:
\begin{enumerate}
    \item Uncertainty quantification in FFF product bond quality
    \item Probabilistic sensitivity analysis
    \item Surrogate modeling of physics model discrepancy
    \item Process parameter optimization under uncertainty
    \item Physical experiments for model calibration and validation
\end{enumerate}
The following subsections describe these steps in detail. 
\subsection{Uncertainty quantification in FFF}\label{Sec:UQinFDM}
In this subsection, several uncertainty sources in FFF are identified and methods to quantify them are discussed.

\subsubsection{Uncertainty sources} \label{Sec:UQ}
The heat transfer and sintering neck growth models used in this paper have their own model inputs, parameters, and errors. Some of these model inputs, parameters, and errors are deterministic while others are uncertain. The uncertainty sources can be aleatory (natural variability) or epistemic (lack of knowledge). The input values used in the heat transfer model (the thickness, width and length of filaments, printer nozzle temperature and extrusion speed) may not be the same as the actual value, thus introducing uncertainty regarding the input to the bond length model. Of the above parameters, the printer nozzle temperature is assumed to vary across printed parts. The temperature of the filaments immediately after being extruded was found to be significantly lower than the specified printer nozzle temperature. This variation in the temperature of the filament as it leaves the nozzle tip is included in the heat transfer model. Moreover, the heat transfer model is also affected by model parameter uncertainty, which is considered epistemic uncertainty (i.e., they have fixed values which are unknown), such as density, specific heat capacity, convective heat transfer coefficient, and fractions of filament perimeter that is in contact with another filament or with the build plate. Thus, the uncertainty in the printer nozzle temperature and model parameters of the heat transfer model introduces uncertainty in the temperature history of the interfaces. The uncertainty in the output of the heat transfer model further propagates to the output quantity of interest through the sintering neck growth model. In addition, the uncertainty regarding the model parameters of the sintering model (such as surface tension, and material viscosity) introduce additional uncertainties in the quantity of interest. Both models also have errors since they have various assumptions, and are not perfect representations of the actual physics. Thus, the bond length predictions have uncertainty due to the propagation of the effects of these different uncertainty sources. The unknown model parameters and errors can be estimated using the experimental data, which contain measurement noise/observation error, through Bayesian model calibration.

\subsubsection{Model calibration under uncertainty}\label{Sec:ModelCalibration}
\begin{figure}[!htb]
    \centering
     \includegraphics[width=.5\columnwidth,keepaspectratio]{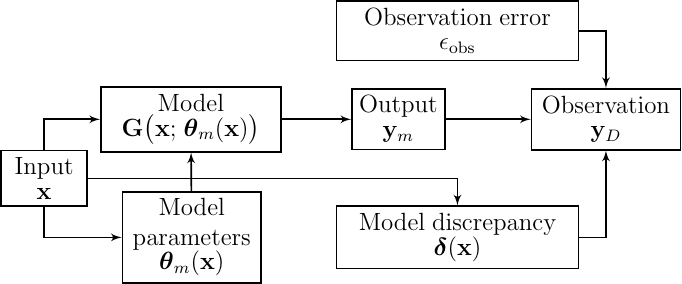}
	\caption{Relating model output to observation data}
    \label{fig:KOH_Framework}
\end{figure}

Consider a single physical quantity of interest $\mathbf{y}$ predicted by a physics model that maps input variables $\mathbf{x}$ and model parameters $\boldsymbol{\theta_m}$ to the numerical model output $\mathbf{y}_{m}$:
\begin{equation}
\mathbf{y}_{m} = \mathbf{G\big(\mathbf{x};}\ \boldsymbol{\theta}_m(\mathbf{x})\big).
\end{equation}
Let $n_{D}$ be the number of collected observation data $\mathbf{y}_D$ from experiments with input variables $\mathbf{x}^{(1)}, ..., \mathbf{x}^{(n_{D})}$, where $\mathbf{x}^{(i)}$ are the input variables for the $i$-th experiment. The difference between observations $\mathbf{y}_D$ and the true response of the system $\mathbf{y}_{\rm true}$ is attributed to observation error $\epsilon_{\rm obs}$, which is often treated as a zero-mean Gaussian random variable with variance $\sigma_{\rm obs}^2$. As discussed in Section~\ref{Sec:UQ}, the model predictions are affected by model errors due to missing physics or approximations. Therefore, a model discrepancy term $\boldsymbol\delta(\mathbf{x})$ as a function of model inputs (one of the main features of the Bayesian calibration framework developed by Kennedy and O'Hagan~\cite{kennedy2001bayesian}) can be introduced as shown in Fig.~\ref{fig:KOH_Framework} to capture the disagreement between $\mathbf{y}_{\rm true}$ and $\mathbf{y}_{m}$. Thus, the true system response can be described as
\begin{equation}
    \mathbf{y}_{\rm true}(\mathbf{x}) = \mathbf{y}_{D}(\mathbf{x}) + \epsilon_{\rm obs}(\mathbf{x})= \mathbf{y}_{m}(\mathbf{x}) + \boldsymbol\delta(\mathbf{x})=\mathbf{G\big(\mathbf{x};}\ \boldsymbol{\theta}_m(\mathbf{x})\big) + \boldsymbol\delta(\mathbf{x}).
    \label{eq:true_response}
\end{equation}

Input variables $\mathbf{x}$ are measurable quantities and chosen by the experimenter. These can be considered deterministic or stochastic with known probability distributions due to natural variability (aleatory) or measurement error. Whereas, model parameters are uncertain due to lack of knowledge (epistemic) since $\boldsymbol{\theta}_m$ take some unknown deterministic values during the experiment. The purpose of Bayesian model calibration is to use observation data $\mathbf{y}_D$ to estimate the posterior distributions of $\boldsymbol{\theta}_m$ and other unknown quantities such as parameters of observation error and the discrepancy term. Kennedy and O'Hagan~\cite{kennedy2001bayesian} employ a probabilistic relationship between the predictions and observations, which incorporates both model parameters and a discrepancy function. Note that the discrepancy function is not observable from the observation data, since the true values of model parameters are unknown. The model discrepancy function is treated as a Gaussian process (GP). The hyperparameters of the GP (including the coefficients of the trend function) can be estimated along with physics model parameters using a Bayesian approach. However, in the presence of insufficient amount of experimental data and non-informative prior knowledge about the uncertainty sources in the engineering system, it may be difficult to distinguish between the effects of the model parameters and model discrepancy; this problem is referred to as non-identifiability~\cite{arendt2012quantification,ling2014selection} when the number of parameters and hyperparameters that need to be estimated becomes large when the model discrepancy term is treated as a GP.

Two strategies are pursued here for improving the identifiability:~(1) performing sensitivity analysis to identify the most important physics model parameters as described in Section~\ref{Sec:Sensitivity}, and~(2) ignoring the discrepancy term during the calibration step and building a surrogate model for the discrepancy (i.e., the difference between the calibrated model prediction and actual system response), as described in Section~\ref{Sec:GP}.

First, the physics model parameters that have the most significant contribution to the uncertainty in the model output during the entire printing process are calibrated without including the model discrepancy term. The joint posterior distribution of $\boldsymbol{\theta}_m$ can be computed using Bayes' theorem:
\begin{equation}
\pi(\boldsymbol{\theta}_m | \mathbf{y}_D) = \frac{L( \mathbf{y}_D | \boldsymbol{\theta}_m; \mathbf{x}^{(1)}, ..., \mathbf{x}^{(n_{D})}) \pi(\boldsymbol{\theta}_m)}{\int_{\Omega_{\boldsymbol{\theta}_m}} L( \mathbf{y}_D | \boldsymbol{\theta}_m; \mathbf{x}^{(1)}, ..., \mathbf{x}^{(n_{D})}) \pi(\boldsymbol{\theta}_m)d\boldsymbol{\theta}_m}
\label{eq:BayesModelCalib}
\end{equation}
where $\Omega_{\boldsymbol{\theta}_m}$ is the domain of $\boldsymbol{\theta}_m$, $\pi(\boldsymbol{\theta}_m)$ is the joint prior distribution of $\boldsymbol{\theta}_m$, $L( \mathbf{y}_D | \boldsymbol{\theta}_m; \mathbf{x}^{(1)}, ..., \mathbf{x}^{(n_{D})})$ is the likelihood function, and the observation data is denoted as $\mathbf{y}_D=[\mathbf{y}_{D}^{(1)}, ..., \mathbf{y}_{D}^{(n_{D})}]\in \mathbb{R}^{n_D}$.

Bayesian model calibration is often performed using Markov chain Monte Carlo (MCMC) sampling algorithms (such as Metropolis-Hastings~\cite{hastings1970monte}, Gibbs~\cite{casella1992explaining}, or slice sampling~\cite{neal2003slice}) since the integral in the denominator of Eq.~\eqref{eq:BayesModelCalib} makes numerical integration intractable for increasing dimension of calibration quantities~\cite{novak2009approximation}. The Metropolis-Hastings algorithm is used in this paper. 

Next, a GP surrogate model is constructed as discussed in Section~\ref{Sec:GP} for the model discrepancy. A set of additional experiments can be conducted to obtain training data for building a surrogate model such as GP model in order to estimate the model discrepancy at any input value $\mathbf{x}$. (Note that two sets of experimental data are used: the first set for calibrating the physics model parameters and the second set for building the surrogate model of model discrepancy. This two-step approach was possible in this study since the experiments were inexpensive and fast; if it is not possible to conduct two sets of experiments, then simultaneous calibration of physics model parameters and GP hyperparameters will need to be pursued, with appropriate assumptions and sensitivity analysis to reduce the number of calibration quantities and achieve identifiability). The training data of the GP model for the model discrepancy can be evaluated for different input values of experimental tests and realizations of observation errors by comparing model predictions against experimental data $\mathbf{y}_{D}(\mathbf{x})$
\begin{equation}
    \boldsymbol{\delta}(\mathbf{x})  = \mathbf{y}_{D}(\mathbf{x}) + \epsilon_{\rm obs} - \mathbf{y}_{m}(\mathbf{x}).
    \label{eq:modelFormError}
\end{equation}

The GP model ($\delta_{GP}(\mathbf{x})$) for the model discrepancy captures the combined contribution of model form and measurement error for a given bond length. Thus, the corrected prediction of the physics-based model $\mathbf{y}_{\rm pred}$ can be written as
\begin{equation}
    \mathbf{y}_{\rm pred}(\mathbf{x}) =  \mathbf{y}_{m}(\mathbf{x}) + \delta_{GP}(\mathbf{x}).
    \label{eq:pred}
\end{equation}

\begin{figure*}[!htb]
\hfill
\begin{subfigure}{.45\textwidth}
  \includegraphics[width=\columnwidth,keepaspectratio]{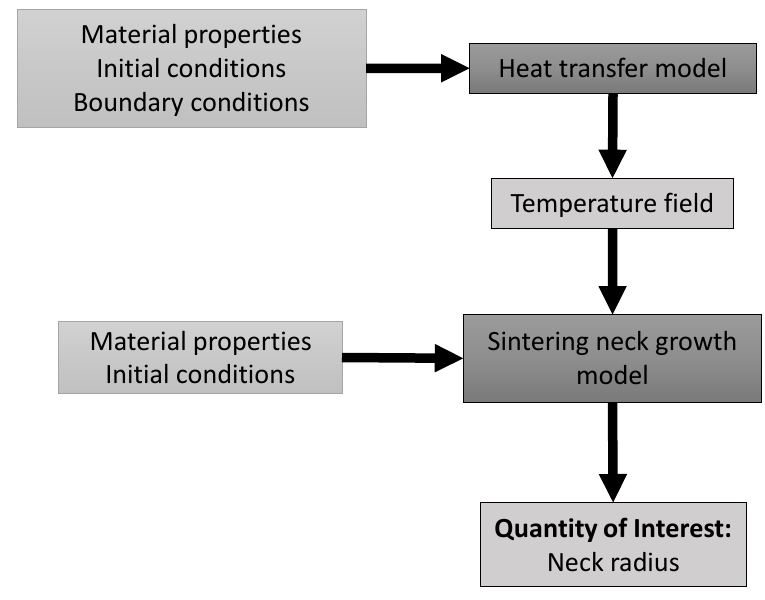}
  \caption{}
\end{subfigure}
\hfill
\begin{subfigure}{.45\textwidth}
  \includegraphics[width=\columnwidth,keepaspectratio]{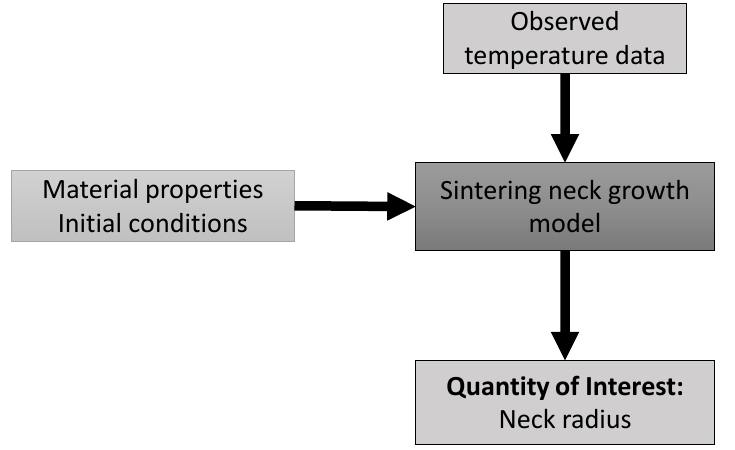}
  \caption{}
\end{subfigure}
\caption{Flowchart of the simulation models for~(a) case A and~(b) case B}
\label{fig:CaseAandB}
\end{figure*}

Two cases with different inputs and model functions are considered in this paper, namely cases A and B. The inputs to the sintering neck growth model, temperature profiles of the extruded filaments, are predicted using the heat transfer model in case A, whereas, measured temperature data from experiments are used as the inputs to the neck growth model in case B. $\mathbf{G\big(\mathbf{x};}\ \boldsymbol{\theta}_m(\mathbf{x})\big)$ represents~(a) the coupled physics-based heat transfer and sintering neck growth models for case A and~(b) sintering neck growth model, which can be evaluated using a numerical technique such as 4th order Runge-Kutta method, for case B~(see~Fig.~\ref{fig:CaseAandB}). In these cases, the physics model is inexpensive to evaluate, thus a surrogate model has not been built for the model $\mathbf{G\big(\mathbf{x};}\ \boldsymbol{\theta}_m(\mathbf{x})\big)$.
\subsection{Probabilistic sensitivity analysis}\label{Sec:Sensitivity}
A probabilistic sensitivity analysis, commonly referred to as global sensitivity analysis (GSA) is used to assess the relative contribution of each uncertainty source towards the uncertainty of the model output (bond length in this case). Model inputs or parameters with negligible contribution can be fixed at their mean values in order to reduce the number of stochastic variables. Variance-based GSA, using Sobol' indices, is adopted in this paper, as briefly described below.

\subsubsection{Variance-based global sensitivity analysis}
Consider a real integrable deterministic one-to-one system response function $\mathrm{Y}=f(\mathbf{X})$, where $\mathbf{X}=\{\mathrm{X}_1, ..., \mathrm{X}_k\}$ are mutually independent model inputs, $f(\boldsymbol{\cdot})$ is the computational model and $\mathrm{Y}$ is the model output. As shown by Sobol~\cite{sobol2001global}, the variance of $\mathrm{Y}$ can be decomposed as
\begin{align}
    V(\mathrm{y}) = & \sum_i^k V_i + \sum_{i_1}^k\sum_{i_2=i_1+1}^k V_{i_1i_2} + \sum_{i_1}^k\sum_{i_2=i_1+1}^k\sum_{i_3=i_2+1}^k V_{i_1i_2i_3} + ... + V_{12...k}
    \label{eq:SobolVar}
\end{align}
where $V_i$ is the variance of $\mathrm{Y}$ due to $\mathrm{X_i}$ alone, and $V_{i_1...i_p} (p\geq2)$ indicates the variance of $\mathrm{Y}$ due to $\{\mathrm{X_{i_1}},...,\mathrm{X_{i_p}}\}$.

The Sobol indices are defined by dividing both sides of Eq.~\eqref{eq:SobolVar} with $V(\mathrm{Y})$
\begin{align}
    1= & \sum_i^k S_i + \sum_{i_1}^k\sum_{i_2=i_1+1}^k S_{i_1i_2} + \sum_{i_1}^k\sum_{i_2=i_1+1}^k\sum_{i_3=i_2+1}^k S_{i_1i_2i_3} + ... + S_{12...k}
    \label{eq:SobolIndex}
\end{align}
where $S_i$ is the first-order or main effects index that assesses the contribution of $\mathrm{X_i}$ individually to the variance of the output $\mathrm{Y}$ without considering interactions with other inputs. The higher-order indices $S_{i_1...i_p} (p\geq2)$ in Eq.~\eqref{eq:SobolIndex} measure the contributions of the interactions among $\{\mathrm{X_{i_1}},...,\mathrm{X_{i_p}}\}$.

In other words, the evaluation of $S_i$ is as follows:
\begin{equation}
    S_i = \frac{V_i}{V(\mathrm{Y})} = \frac{V_{X_i}(E_{\mathbf{X}_{-i}}(\mathrm{Y}|\mathrm{X}_i))}{V(\mathrm{Y})}
    \label{eq:FirstIndex}
\end{equation}
where $\mathbf{X}_{-i}$ are all the model inputs other than $\mathrm{X}_i$.

The overall contribution of $\mathrm{X_i}$ considering an individual input and its interactions with all other inputs is measured by the total effects index $S_i^T$:
\begin{equation}
    S_i^T = 1 - \frac{V_{-i}}{V(\mathrm{Y})} = \frac{V_{\mathbf{X}_{-i}}(E_{\mathrm{x}_i}(\mathrm{Y}|\mathbf{X}_{-i}))}{V(\mathrm{Y})}.
    \label{eq:TotalIndex}
\end{equation}
The calculation of first-order and total effects indices requires a deterministic function. Furthermore, the total effects index $S_i^T$ is only meaningful for uncorrelated model inputs~\cite{saltelli2002relative}. Whereas, the first-order index $S_i$ can be calculated for both correlated and uncorrelated model inputs~\cite{saltelli2002relative}.

The computation of $S_i$ analytically is nontrivial since $E_{\mathbf{X}_{-i}}(\boldsymbol{\cdot})$ requires multi-dimensional integrals. Using Monte Carlo simulation (MCS) to measure $S_i$ is also expensive because calculation of the numerator of $S_i$ requires a double-loop MCS,  where the inner loop $E_{\mathbf{X}_{-i}}(\mathrm{Y}|\mathrm{X}_i)$ computes the mean value of $\mathrm{Y}$ using $n_1$ random samples of $\mathbf{X}_{-i}$ and the outer loop computes $V_{X_i}(E_{\mathbf{X}_{-i}}(\mathrm{Y}|\mathrm{X}_i))$ by iterating the inner loop $n_2$ times for different values of $\mathrm{X}_i$. Moreover, the computation of $V(\mathrm{Y})$ requires additional $n_3$ MCS iterations. The total computational cost of the first-order and total effect indices (i.e. the number of function evaluations, $N_f$) is approximately $N_f = kN^2+N$, where $N=n_1=n_2=n_3$. Thus, the calculation of indices become unaffordable for a large number of inputs and expensive models since the required number of random samples are of the order greater than 1000 in many practical applications.

The sampling-based method modularized global sensitivity analysis (MGSA) proposed by Li and Mahadevan~\cite{li2016efficient}, which has a computational cost that is not proportional to the model input dimension, can directly estimate the first-order Sobol' index from Monte Carlo samples with a single-loop instead of the double-loop MCS. The first-order Sobol' index can be computed by dividing the input variable $X_i$ into equally probable intervals $\boldsymbol{\phi}={\phi^1,...,\phi^M}$:
\begin{equation}
    S_i = 1 - \frac{E_{\phi}(V_{\phi^p}(Y))}{V(Y)},\ p=1,...,M
\end{equation}
where $V_{\phi^p}(Y)$ is the variance of $Y$ when $X_i$ in the subspace $\phi^p$, $V(Y)$ represents the variance of the system response $Y$. The main advantages of this efficient data-driven method based on the concept of stratified sampling~\cite{iooss2009global} are that the computational cost is not proportional to the number of model inputs and the physics/computational model does not have to be available since the input-output samples are enough to compute the first-order Sobol index. Further advances in this direction based on importance sampling and kernel functions have been reported by DeCarlo et al.~\cite{decarlo2018efficient}.

\subsection{Gaussian process (GP) surrogate modeling}\label{Sec:GP}
In this paper, a GP surrogate model is constructed to estimate the model discrepancy at unobserved inputs. The GP surrogate model properly accounts for statistical dependence between the outputs at different input values, and provides the output at a given input value as a normal distribution with an expected value and a standard deviation.

In the GP model, the response at prediction point \textbf{u}, $G$(\textbf{u}) is described by:
\begin{equation}
G(\mathbf{u})=\mathbf{h}(\mathbf{u})^{T} \boldsymbol{\beta}+Z(\mathbf{u})
\end{equation}
where $\mathbf{h}(\cdot)$ is the trend of the model, $\boldsymbol{\beta}$ is the vector of trend coefficients, and $Z(\cdot)$ is a zero-mean stationary Gaussian process which describes the deviation of the model from the trend. The covariance between the outputs $Z(\cdot)$ of the GP surrogate at points \textbf{a} and \textbf{b} is defined as:
\begin{equation}
\operatorname{Cov}[Z(\mathbf{a}), Z(\mathbf{b})]=\sigma_{Z}^{2} R(\mathbf{a}, \mathbf{b})
\label{eq:Cov}
\end{equation}
where $\sigma_Z^2$ is the process variance and $R(\boldsymbol{\cdot},\boldsymbol{\cdot})$ is the correlation function. A squared exponential function with separated length scale parameters $l_i$ for each input dimension has often been used in the literature:
\begin{equation}
R(\mathbf{a}, \mathbf{b})=\exp \left[-\sum_{i=1}^{\rm M} \frac{\left(a_{i}-b_{i}\right)^{2}}{l_i}\right]
\end{equation}
The outputs of the GP model are the mean prediction $\mu_G(\cdot)$ and the variance of the prediction $\sigma_G^2(\cdot)$, defined as:
\begin{equation}
\mu_{G}(\mathbf{u})=\mathbf{h}(\mathbf{u})^{T} \boldsymbol{\beta}+\mathbf{r}(\mathbf{u})^{T} \mathbf{R}^{-1}(\mathbf{g}-\mathbf{F} \boldsymbol{\beta})
\label{eq: mean}
\end{equation}
\begin{equation}
\sigma_{G}^{2}(\mathbf{u})=\sigma_{Z}^{2}- \mathbf{A} \left[ \begin{array}{cc}{\mathbf{0}} & {\mathbf{F}^{T}} \\ {\mathbf{F}} & {\mathbf{R}}\end{array}\right]^{-1} \mathbf{A}^T
\label{eq: sd}
\end{equation}
where $\mathbf{r}(\mathbf{u})$ is a vector containing the covariance between $\mathbf{u}$ and each of the training points \{$\boldmath{x_1}, \boldmath{x_2}, ..., \boldmath{x_{n}}$\}, $i\in \{1,...,n\}$, $\mathbf{R}$ is an $n \times n $ matrix containing the correlation between each pair of training points, $\mathbf{R}(\boldmath{x_i},\ \boldmath{x_j}) =$ Cov$[Z(\boldmath{x_i}), Z(\boldmath{x_j})]$; $\mathbf{g}$ is the vector of original physics model outputs at each of the training points, $\mathbf{F}$ is a $n\times q$ matrix with rows $\mathbf{h}(\mathbf{u}_i)^T$, and $\mathbf{A}$ = $\left[\mathbf{h}(\mathbf{u})^{T} \ \mathbf{r}(\mathbf{u})^{T}\right]$.

The trained GP surrogate model is used to correct the model prediction, which feeds into the process parameter optimization calculations as presented in Section~\ref{Sec:Optim}.

\subsection{Process optimization under uncertainty}\label{Sec:Optim}
The optimal design point that satisfies design criteria and a specified level of reliability is of great interest in many engineering applications. In this paper, the focus is on selecting the optimum values of process parameters that maximize the bond lengths between filaments and between layers. The AM process optimization under uncertainty can be pursued in two directions:~(1) reliability-based design optimization (RBDO)~\cite{du2004sequential}, and~(2) robust design optimization (RDO)~\cite{zaman2011robustness}. In RBDO, the decision variables are optimized to either maximize or achieve a desired target level of reliability (i.e., probability of satisfying a desired threshold of performance or quality). In RDO, the decision variables are optimized such that the variability of the objective function is minimized, and the constraints are satisfied within specified uncertainty bounds. The approach of robustness-based design optimization (RDO) is used to maximize the overall bond quality. The robustness of the objective function can be achieved by simultaneously optimizing the mean and variance; thus, this is a bi-objective problem. Monte Carlo sampling is used to compute the mean and variance of the objective function (the mean bond length of a layer) in the probabilistic optimization process. An efficient sampling-based method Latin hypercube sampling is used to simulate the uncertain parameters.

The robust design optimization problem can be formulated as follows:
\begin{equation}
    \begin{aligned}
    & \underset{\mathbf{d}  \in \mathbb{R}^{n_d}}{\text{minimize}}
    & & \big\{E(y_o(\mathbf{d})), V(y_o(\mathbf{d}))\big\}; \\
    & \text{subject to}
    & & \big\{E(y_c(\mathbf{d})), V(y_c(\mathbf{d}))\big\}\leq 0,\ c=1,2,...,n_c; \\
    & 
    & & d_{i,\rm lb} \leq d_i \leq d_{i,\rm ub},\ i=1,2,...,n_v,
    \end{aligned}
\label{eq:RDOframework}
\end{equation}
where the objective function and constraints are expressed as functions of expectation $E(\cdot)$ and variance $V(\cdot)$ of objective function $y_o$ and constraints $y_c$, respectively; the vector of design variables $\mathbf{d}$ can be either deterministic parameters or mean/standard deviation of the design parameters, $n_c$ and $n_v$ are the number of constraints and design variables, respectively. The $i$-th design variable $d_i$ is constrained by its lower bound $d_{i,\rm lb}$ and upper bound $d_{i,\rm ub}$. The RDO problem becomes bi-objective by adding the variance of the performance function to the expected value of the objective function, and a weighted sum method can be used to assign proportional weights for the aggregation of the two objectives according to their importance~\cite{chen1996procedure}. The aggregation formulation is given by Eq.~\eqref{eq:bi-obj}.
\begin{equation}
    \begin{aligned}
    & \underset{\mathbf{x}  \in \mathbb{R}^{n_x}}{\text{minimize}}
    & & w_1 E(y_o(\mathbf{d})) + w_2 V(y_o(\mathbf{d})); \\
    & \text{subject to}
    & & \mathbf{h}(\mathbf{d})=0,\ \\
    &
    & & \mathbf{g}(\mathbf{d})\leq 0,\ \\
    & 
    & & d_{i,\rm lb} \leq d_i \leq d_{i,\rm ub},\ i=1,2,...,n_v,
    \end{aligned}
\label{eq:bi-obj}
\end{equation}
where $w_1, w_2 >0$ are the weighting coefficients representing the relative importance of each objective function, and $\mathbf{h}(\mathbf{d})$ and $\mathbf{g}(\mathbf{d})$ are the vectors of equality and inequality constraints respectively.

The \emph{nested} bi-objective robustness-based design optimization (RDO) problem can be converted into a single objective formulation, using a weighted sum approach, as:
\begin{equation}
    \begin{aligned}
    & \underset{\mathbf{x}  \in \mathbb{R}^{n_x}}{\text{minimize}}
    & & w_1\mathrm{F}_{1} + w_2\mathrm{F}_{2} \\
    & \text{subject to}
    & & \mathbf{x}_{\rm lb} \leq \mathbf{x} \leq \mathbf{x}_{\rm ub} \\
    \end{aligned}
    \label{eq:RDO}
\end{equation}
where weighting coefficients $w_1, w_2 >0$ represent the relative importance of two objectives. $\mathrm{F}_{1}=-\mathrm{\mu}_{\boldsymbol{\mu}_{\rm BL,i}}+\mathrm{\sigma}_{\boldsymbol{\mu}_{\rm{BL},i}}$ represents the mean and standard deviation of the average bond length predictions $\mu_{\rm{BL}}$ for layer $i$, and $\mathrm{F}_{2}=\mathrm{\mu}_{\boldsymbol{\sigma}_{\rm{BL},i}}+\mathrm{\sigma}_{\boldsymbol{\sigma}_{\rm{BL},i}}$  represents the mean and standard deviation of the standard deviation of the bond length predictions $\sigma_{\rm BL}$ for layer $i$, $i\in \{1,...,M\}$. $M$ is the total number of layers, $\mathbf{x}$ are the design variables (printer nozzle temperature and printer extrusion speed for each layer $i$), and $\mathbf{x}_{\rm lb} \leq \mathbf{x} \leq \mathbf{x}_{\rm ub}$ represents the lower and upper bounds for the design variables. The weighted sum approach is a convex combination of two different objectives, $\mathrm{F}_{1}$ and $\mathrm{F}_{2}$. The solution of the optimization problem approximates the Pareto front by changing the weights of each objective. The Pareto front maps the relation between these two objective functions. The negative of the mean value of mean bond lengths at each layer is minimized while minimizing the deviation of mean bond lengths at each layer with the use of function $\mathrm{F}_{1}$. The function $\mathrm{F}_{2}$, which is a convex combination of the mean and standard deviation of the deviation of bond lengths at each layer, is minimized simultaneously with $\mathrm{F}_{1}$. In other words, the overall bond quality (the mean value of bond length for a part) is maximized, while minimizing the variations in the quantity of interest (bond lengths between filaments) using the functions $\mathrm{F}_{1}$ and $\mathrm{F}_{2}$ respectively.

\subsection{Experimental work}
\begin{figure}[!htb]
  \centering
    \begin{subfigure}{.45\textwidth}
      \centering
      \includegraphics[width=\columnwidth,keepaspectratio]{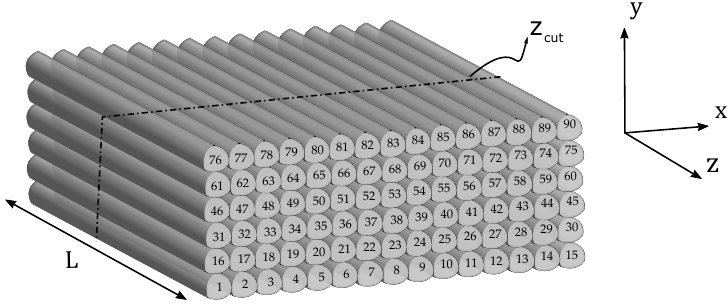}
      \caption{}
      \label{fig:DeposSeqSequentiala}
    \end{subfigure}
    \begin{subfigure}{.45\textwidth}
      \centering
      \includegraphics[width=\columnwidth,keepaspectratio]{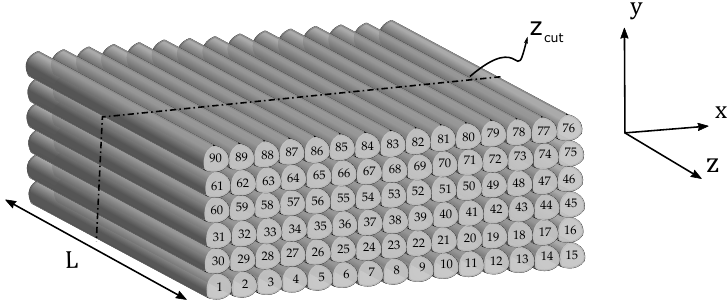}
      \caption{}
      \label{fig:DeposSeqSequentialb}
    \end{subfigure}
\caption{Deposition sequence of unidirectional 90 filaments:~(a) from left to right for all the layers and~(b) from left to right in odd numbered layers and from right to left in even numbered layers}
\label{fig:DeposSeqSequential}
\end{figure}

A commercial material, Ultimaker Black ABS, was used in the experiments. A unidirectional and aligned building strategy was adopted at a specified printer nozzle temperature and extrusion speed. Two different options for the deposition sequence of the filaments were considered, as shown in Fig.~\ref{fig:DeposSeqSequential}. In Fig.~\ref{fig:DeposSeqSequentiala}, the filaments are sequenced from left to right in all the layers. In Fig.~\ref{fig:DeposSeqSequentialb}, the filaments are sequenced from left to right in odd numbered layers and from right to left in even numbered layers. The filament numbers in the two figures correspond to the two deposition sequence options.

Multiple rectangular-shaped specimens were produced with the same geometry but different combinations of process parameter values. For each specimen, the temperature distribution at the top of each layer during deposition was monitored using an infrared thermography camera. Thermal images were recorded with a specified frequency until all filaments were deposited. The neck growth between the filaments and the total void area of the parts were identified at a specified cross-section with the use of microscopy images processed through the ImageJ software~\cite{schneider2012nih}. The statistical properties of the neck growth along the length of the specimens were constant. Therefore, all specimens were sectioned at the midpoint to analyze the mesostructural feature of interest only at that cross-section.

\section{Numerical results and discussion}\label{Sec:Results}
\begin{figure}[!htb]
\centerline{\includegraphics[width=.5\columnwidth,keepaspectratio]{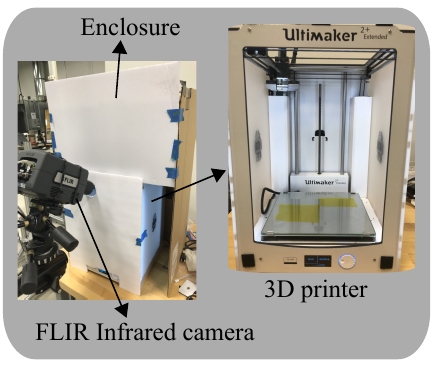}}
\caption{The experimental setup}
\label{fig:ExpSetup}
\end{figure}

The experimental setup used to build rectangular acrylonitrile butadiene styrene (ABS) amorphous polymer specimens of length 35 mm, width 12 mm, and thickness 4.2 mm is shown in Fig.~\ref{fig:ExpSetup}. The specimens were created on an Ultimaker 2 extended+ printer, which is within an enclosure to reduce the part variability; a commercial material, Ultimaker Black ABS, was used. Air flow was not considered because we enclosed the printer to prevent the occurrence of airflow. All parts were printed through a nozzle with 0.8 mm diameter. The build plate temperature was constant and set to 110$^{\circ}$C and the environment temperature is assumed to be 70$^{\circ}$C. The extrusion rate and vertical position of the nozzle were adjusted by the printer to be able to produce each filament with 0.8 mm width and 0.7 mm height.

The surface temperature profiles of extruded filaments were monitored using an infrared thermography camera as shown in Fig.~\ref{fig:ExpSetup}. The extrusion of the next layer prevents the camera from monitoring the temperature profiles of the previous layers. Due to the inability to obtain temperature data of the filaments below the top layer, we could only predict the quality of the intra-layer bonding using temperature profiles of extruded filaments within that layer. Thermal images were recorded with a frequency of 10 Hz until the deposition of all filaments was completed.

All specimens used in this study are produced with unidirectional filaments to enhance the effects of process parameters on the bond quality between adjoining filaments.\ Each filament of the rectangular part is deposited at a specified printer nozzle temperature $T_n$ and extrusion speed $v_p$. The temperature evolution of the interfaces, and the neck growth between the filaments (the mesostructural feature of interest), are predicted at $z_{\rm cut}=L/2$ as shown in Fig.~\ref{fig:DeposSeqSequential}. The process parameters and material properties used in this work are presented in Table~\ref{table:Parameters}. The specific heat capacity and density of the material are calibrated together as a single term $\alpha=\rho C$, where $\rho$ and $C$ are density (kg/m$^3$) and specific heat capacity (J/kg $^{\circ}$C) respectively. The analysis assumes temperature dependent material properties such as material viscosity $\eta$ and surface tension $\Gamma$. The surface tension of ABS P400 at 240$^{\circ}$C is 0.029 N/m as reported by Bellehumeur et al.~\cite{bellehumeur2004modeling} with a temperature dependence $\Delta\Gamma/\Delta T = -\gamma$ N/m $\cdot$ K, where the neck growth model parameter $\gamma=$ 0.00345. The temperature dependent material viscosity $\eta$ is given by $\eta=\eta_r\ \text{exp}[-\beta (T-T_r)]$, where the material viscosity at the reference temperature ($T_r=$ 240$^{\circ}$C) $\eta_r$ is 5100 Pa $\cdot$ s,\ $\beta$ is a model parameter that is selected as 0.056 by Sun et al.~\cite{sun2008effect}, and $T$ is the temperature of the material at a given time instance.
\begin{table*}[!htb]
\small
\caption{Process parameters and material properties}
\label{table:Parameters}
\centering{%
    \begin{tabular*}{\linewidth}{@{\extracolsep{\fill}}lc@{\extracolsep{\fill}}}
    \toprule
    Property & Value\\
    \midrule
            Printer nozzle temperature ($^{\circ}$C) & 240 \\
            Build plate temperature ($^{\circ}$C) & 110 \\
            Printer extrusion speed (m/s) & 0.042 \\
            Filament length (m) & 0.035 \\
            Filament width (m) & 0.0008 \\
            Filament thickness (m) & 0.0007 \\
            Fraction of filament's perimeter for all contacts & 0.15 \\
            Convective heat transfer coefficient (W/m$^2$ $^{\circ}$C) & 86\\
            Conductive heat transfer coefficient between filaments (W/m$^2$ $^{\circ}$C) & 200\\
            Conductive heat transfer coefficient between filament and build plate (W/m$^2$ $^{\circ}$C) & 86\\
            Thermal conductivity (W/m $^{\circ}$C) & 0.15\\
            $\alpha$ (J/m$^3$ $^{\circ}$C) & 1.196 $\times$ 10$^{6}$\\
    \bottomrule
    \end{tabular*}
}%
\end{table*}

\subsection{Prediction of the cooling of filaments}
\begin{figure}[!htb]
\centerline{\includegraphics[width=.5\columnwidth,keepaspectratio]{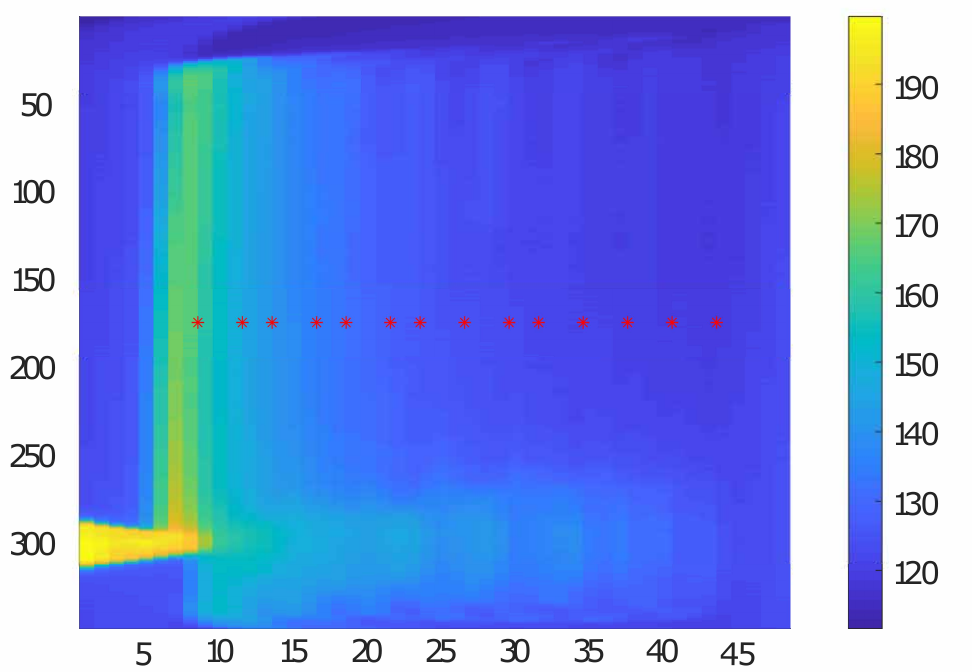}}
	\caption{Top view temperature profile of the first layer}
	\label{fig:ExpTemp24042}
\end{figure}

A typical IR image of the temperature profile for the first layer of a part printed using $(T_n,v_p)=$ (240$^{\circ}$C, 0.042\ m/s) is shown in Fig.~\ref{fig:ExpTemp24042}. The interface temperature is monitored at corresponding locations between filaments. The experimental temperature profile was used to assess the validity of the heat transfer model in order to be able to predict the neck growth accurately using the heat transfer model predictions. The temperature of the filaments immediately after being extruded onto the build plate or onto another filament was found to be significantly lower (20$^{\circ}$C to 50$^{\circ}$C) than the specified printer nozzle temperature. At the upper temperature limit of the printer the filaments were extruded at temperatures approximately 40-50$^{\circ}$C less than the set nozzle temperature. The enclosure was not a precisely controlled environmental chamber with quantitative measurements of temperature, humidity, and airflow. Therefore, these effects might be influencing the difference between the printer setting and observed temperature. This variation in the temperature of the filament as it leaves the nozzle tip is considered as a bias term in the heat transfer model.
\begin{figure}[!htb]
\begin{center}
\begin{subfigure}{.45\textwidth}
  \centering
  \includegraphics[width=\columnwidth,keepaspectratio]{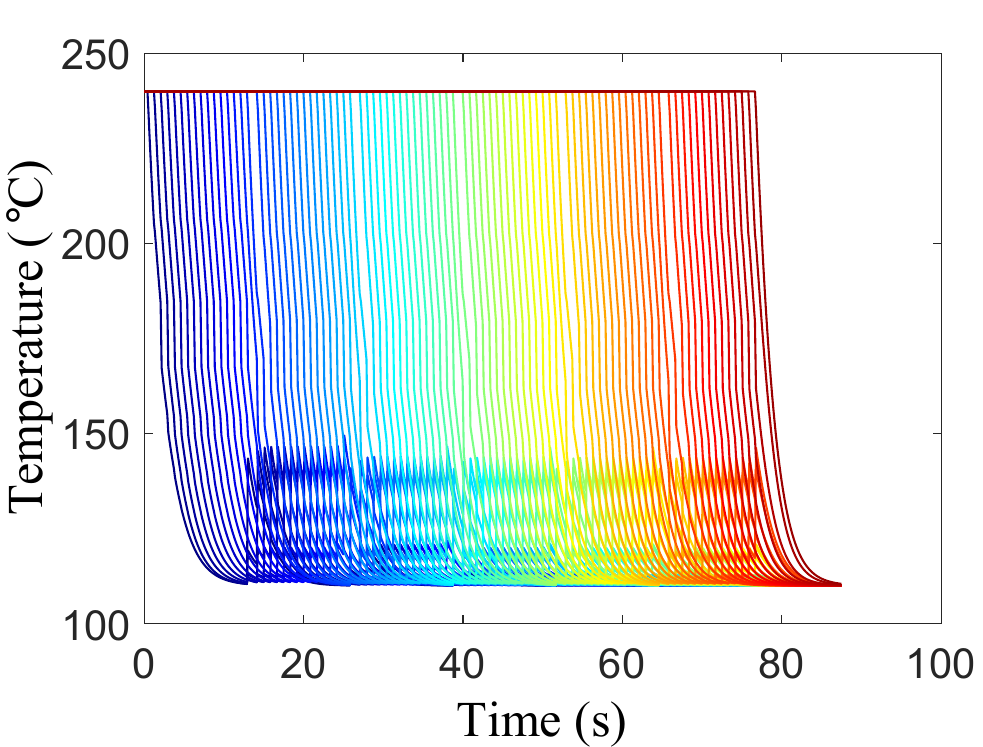}
  \caption{}
  \label{fig:TempProfa}
\end{subfigure}
\begin{subfigure}{.45\textwidth}
  \centering
  \includegraphics[width=\columnwidth,keepaspectratio]{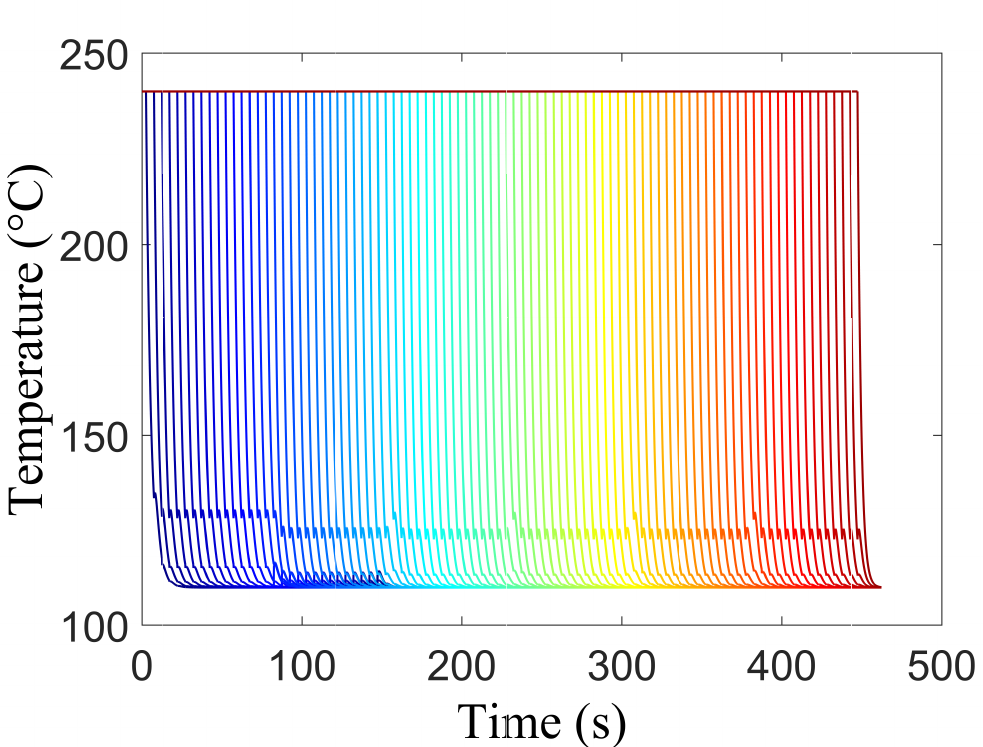}
  \caption{}
  \label{fig:TempProfb}
\end{subfigure}
\end{center}
\caption{Temperature evolution of 90 filaments at $z_{\rm cut}=L/2$~(a) with $L=$ 0.035 m shown in Fig.~\ref{fig:DeposSeqSequentiala} and~(b) with $L=$ 0.21 m shown in Fig.~\ref{fig:DeposSeqSequentialb} along deposition time}
    \label{fig:TempProf}
\end{figure}

The temperature evolution of all the filaments illustrated in Fig.~\ref{fig:DeposSeqSequentiala} and Fig.~\ref{fig:DeposSeqSequentialb} at $z_{\rm cut}=$ 0.0175 m along deposition time is shown in Fig.~\ref{fig:TempProfa} and Fig.~\ref{fig:TempProfb} respectively. The length to diameter ratio of filaments has a significant effect on the cooling process. The time it takes for the printer to extrude a single filament increases as the lengths of the filaments get longer. This results in a faster cooling process, and consequently a smaller amount of heat transfer between each filament. Moreover, extruding each layer's first filament at the same $x$-coordinate results in a more homogeneous part quality as the temperature difference between the filaments extruded on top of the filaments below is approximately the same. Whereas, in Fig.~\ref{fig:DeposSeqSequentialb}, the temperature difference between the 1st and 30th filaments is much greater than 15th and 16th filaments, resulting in a staggered temperature evolution and part quality. Therefore, the build strategy shown in Fig.~\ref{fig:DeposSeqSequentiala} is used for further analysis.
\begin{figure}[!htb]
\centerline{\includegraphics[width=.55\columnwidth,keepaspectratio]{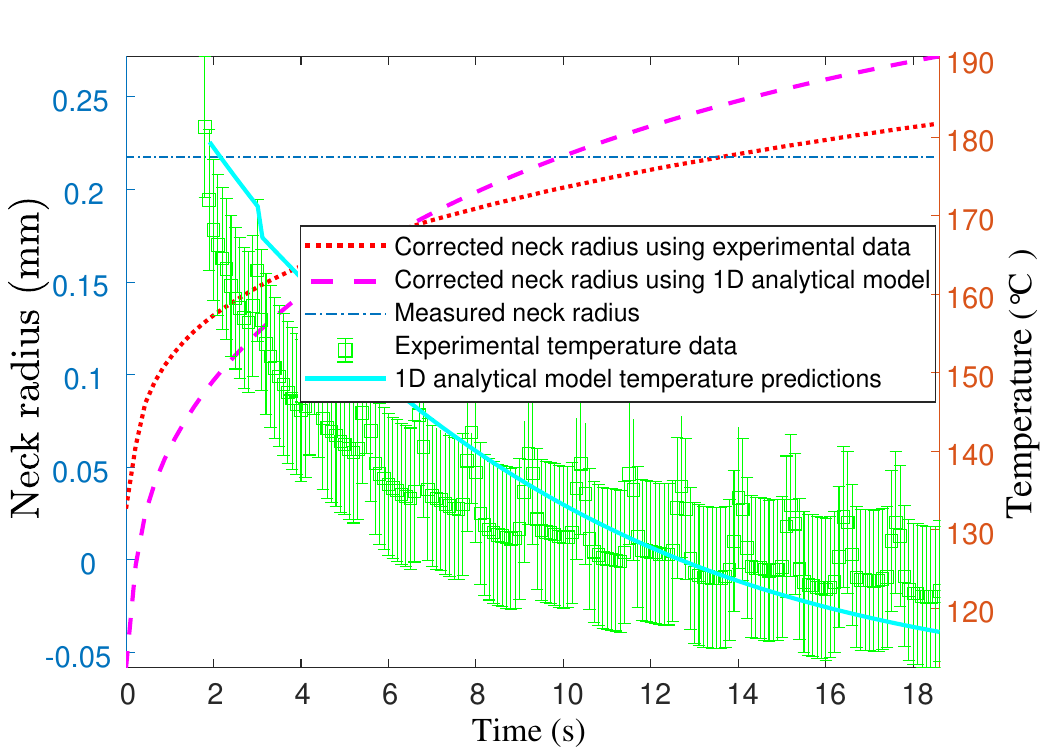}}
	\caption{Experimental temperature profiles compared with model predictions for the interface between filaments 1 and 2 at $z_{\rm cut}=$ 0.0175 m and the neck growth predictions using the heat transfer model predictions (case A) and experimental temperature data (case B)}
	\label{fig:CorrectedPred}
\end{figure}

The Newtonian sintering model is coupled with the heat transfer model to predict the bond lengths between filaments. The one-dimensional transient heat transfer model predictions, and observed temperature data for the interface between filaments 1 and 2 are compared in Fig.~\ref{fig:CorrectedPred}. The model predictions are in general agreement with the measured data. The measured temperature data and model predictions show a similar trend in the initial stage due to enhanced convection at higher temperatures, but at temperatures below 130$^{\circ}$C the model prediction deviates away from the measurement data. However, the inaccuracy of the model at lower temperatures is not relevant for neck growth predictions since the neck growth process occurs at higher temperatures~\cite{bellehumeur2004modeling}.

The neck radius predictions for each case are corrected with the model discrepancy (estimated by the surrogate model) at given input values. The measured neck radius and the corrected neck growth predictions corresponding to case A (heat transfer model predictions as the input to the sintering neck growth model) and case B (observed temperature profile as the input to the sintering neck growth model) are demonstrated in Fig.~\ref{fig:CorrectedPred} for $(T_n,v_p)=$ (240$^{\circ}$C, 0.042\ m/s). The corrected neck radius predictions are in agreement with the measured data around 130$^{\circ}$C since the model parameters are calibrated using the neck radius predictions when the interface temperature is at 130$^{\circ}$C. Thus, case A is used for further analysis when the temperature data is not available.
\begin{figure}[!htb]
\begin{center}
    \begin{subfigure}{.45\textwidth}
      \includegraphics[width=\textwidth,keepaspectratio]{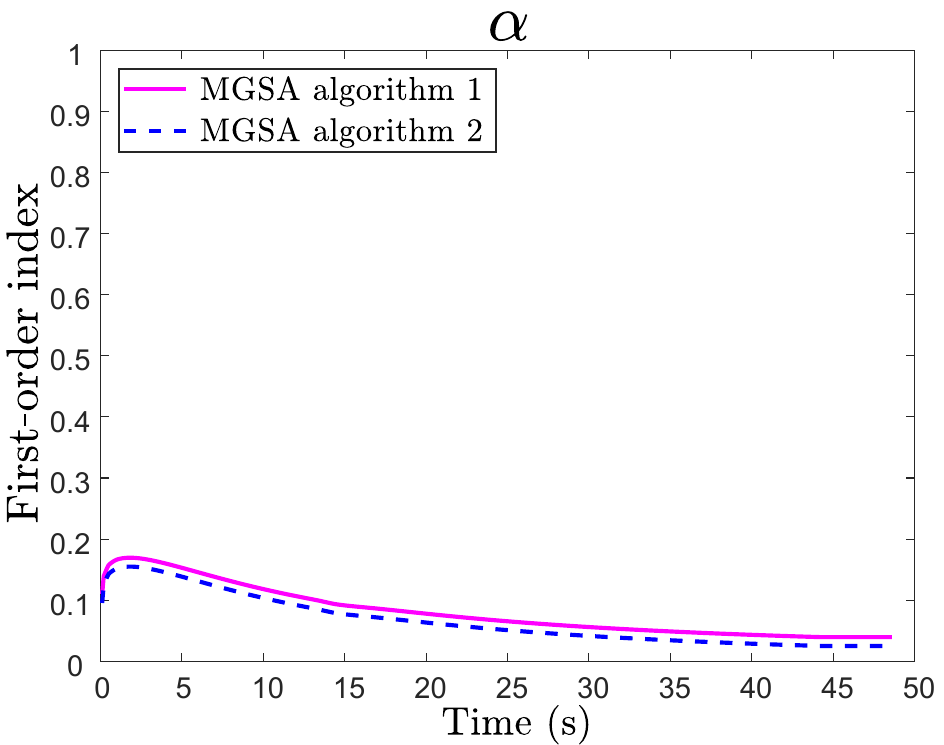}
      \caption{}
    \end{subfigure}
    \begin{subfigure}{.45\textwidth}
      \includegraphics[width=\textwidth,keepaspectratio]{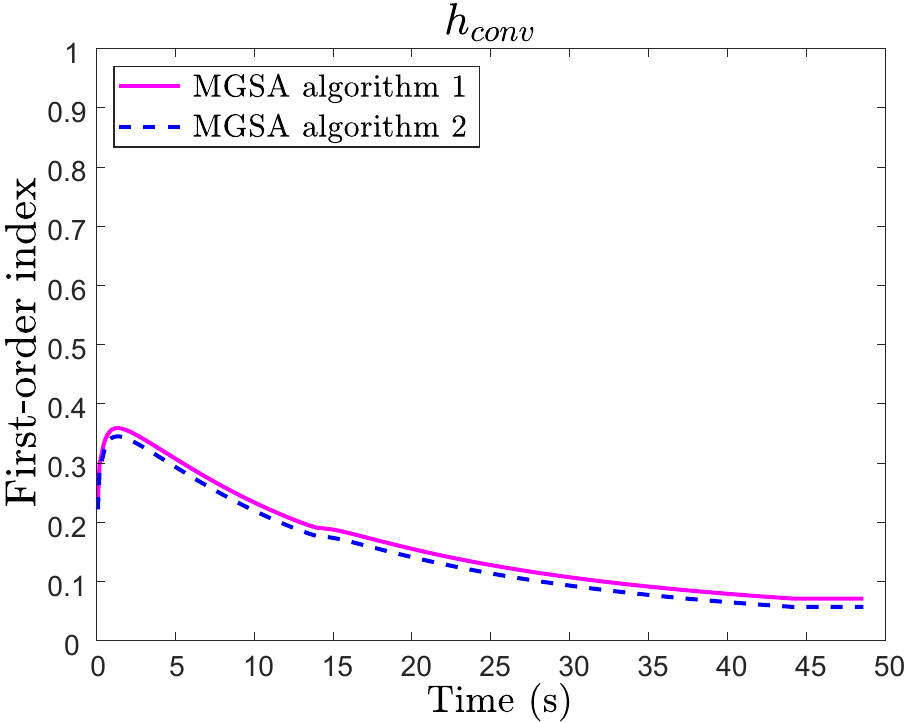}
      \caption{}
    \end{subfigure}
    \\ 
    \begin{subfigure}{.45\textwidth}
      \includegraphics[width=\textwidth,keepaspectratio]{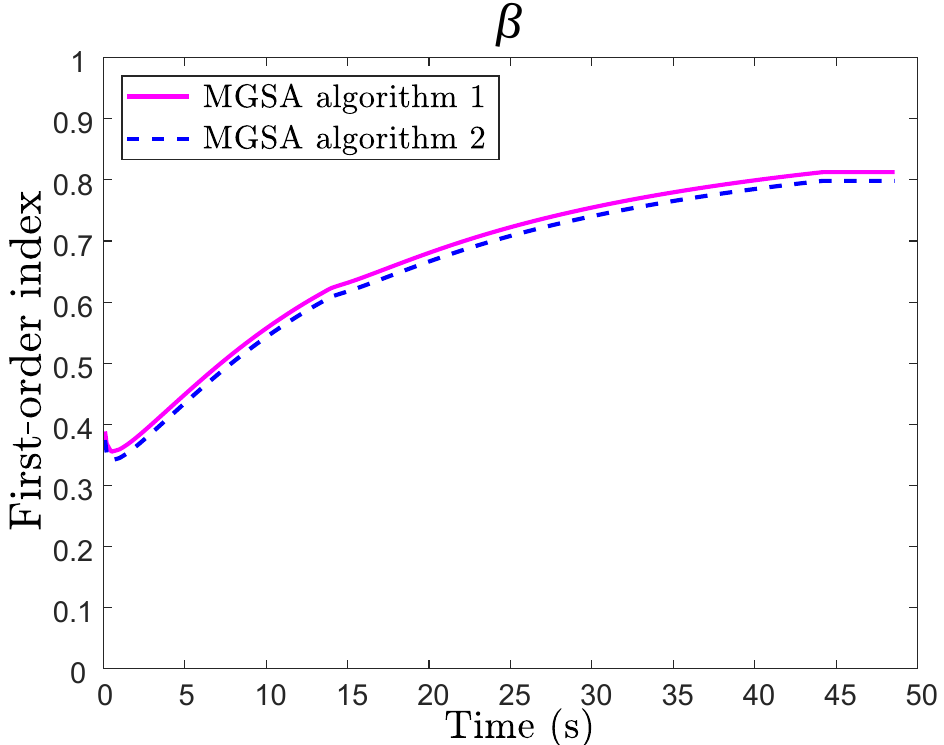}
      \caption{}
      \label{SensitivityIndices_BLModel1c}
    \end{subfigure}
    \begin{subfigure}{.45\textwidth}
      \includegraphics[width=\textwidth,keepaspectratio]{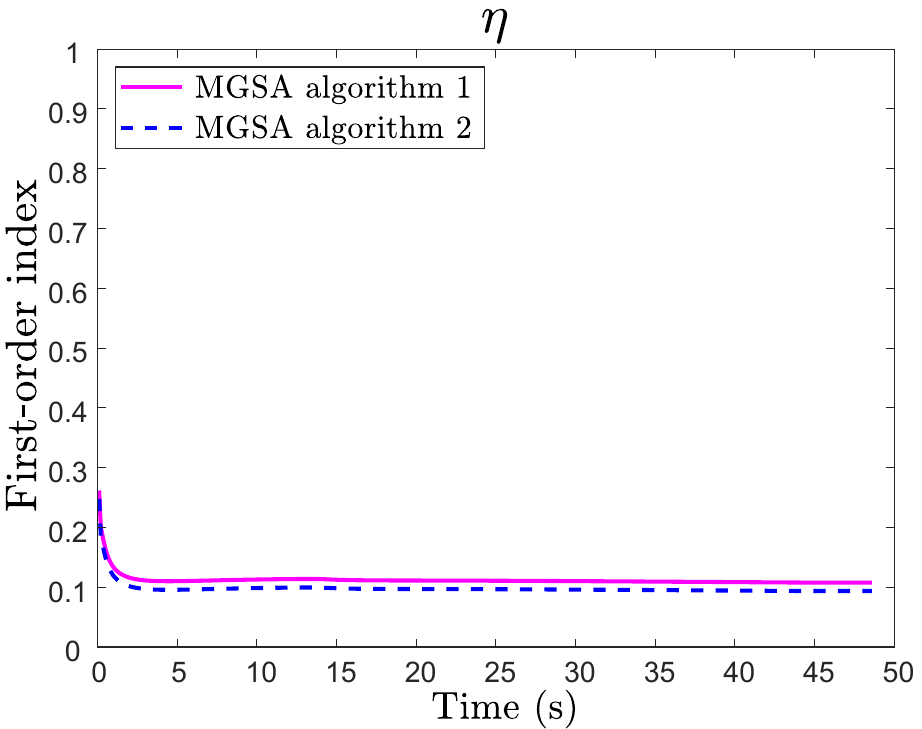}
      \caption{}
    \end{subfigure}
    \end{center}
\caption{Sensitivity indices for the bond length between the 1st and 2nd filaments in the first layer}
\label{SensitivityIndices_BLModel1}
\end{figure}

\subsection{Contribution assessment of each uncertainty source}\label{Sec:GSA}
A sample-based single loop algorithm called MGSA (modularized GSA) proposed by Li and Mahadevan~\cite{li2016efficient} is used to compute the first-order Sobol' indices. The results from MGSA indicate which parameters' individual effect have significant contribution to the uncertainty in the coupled heat transfer and neck growth models. A low first-order index implies that the individual effect of the parameter is insignificant; thus, it can be fixed at its mean value. Thus, GSA provides insights on where to focus resources for improving the AM process.

The random variables in the heat transfer model are $\alpha, h_{conv}$, and $\lambda_i$, $i\in \{1,2,3,4,5 \}$, i.e., the material parameter, convective heat transfer coefficient and fraction of filament's perimeter that is in contact with other filaments or with the build plate. The random variables in the sintering neck growth model are $\Gamma,\ \eta,\ \beta$ , and $\gamma$, i.e., the surface tension, material viscosity values at reference temperature of 240$^{\circ}$C, and model parameters of the temperature dependent surface tension and material viscosity respectively.

\subsubsection{GSA of the bond length model}
\begin{figure*}[!htb]
\begin{center}
    \begin{subfigure}{.45\textwidth}
      \includegraphics[width=\textwidth,keepaspectratio]{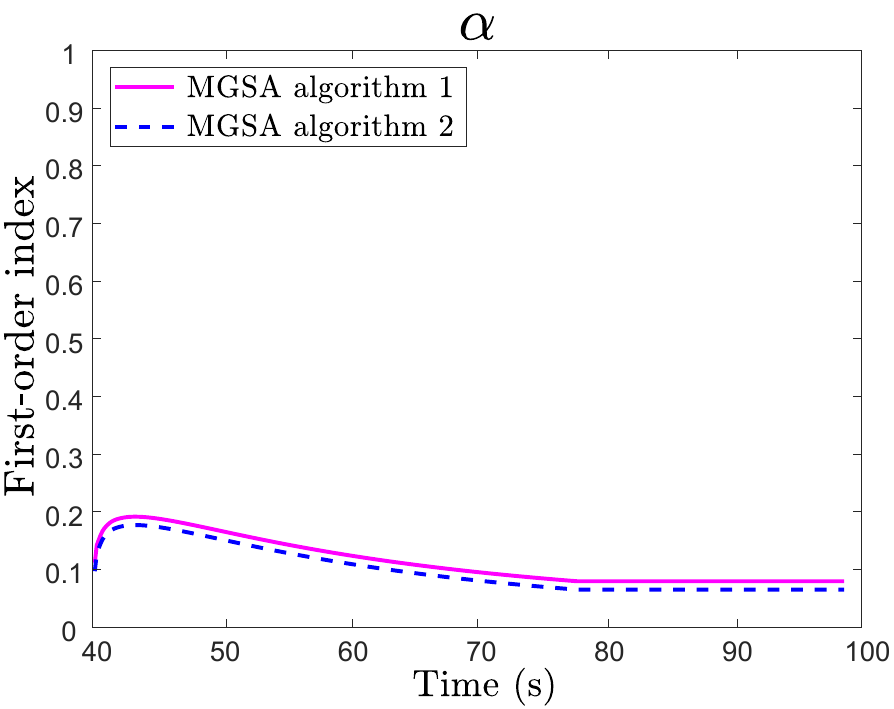}
      \caption{}
    \end{subfigure}
    \begin{subfigure}{.45\textwidth}
      \includegraphics[width=\textwidth,keepaspectratio]{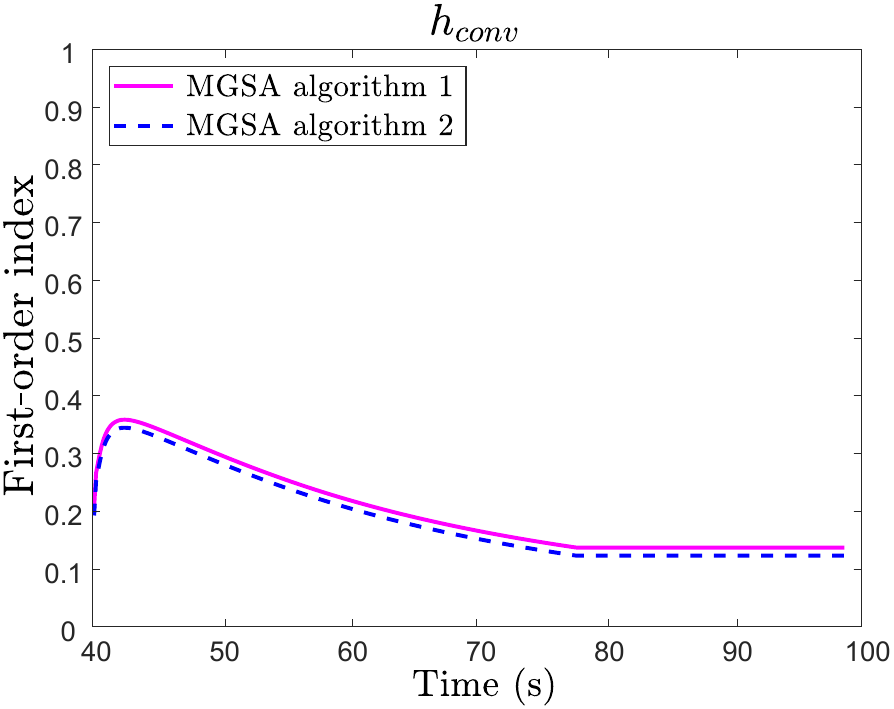}
      \caption{}
    \end{subfigure}
\\ 
    \begin{subfigure}{.45\textwidth}
      \includegraphics[width=\textwidth,keepaspectratio]{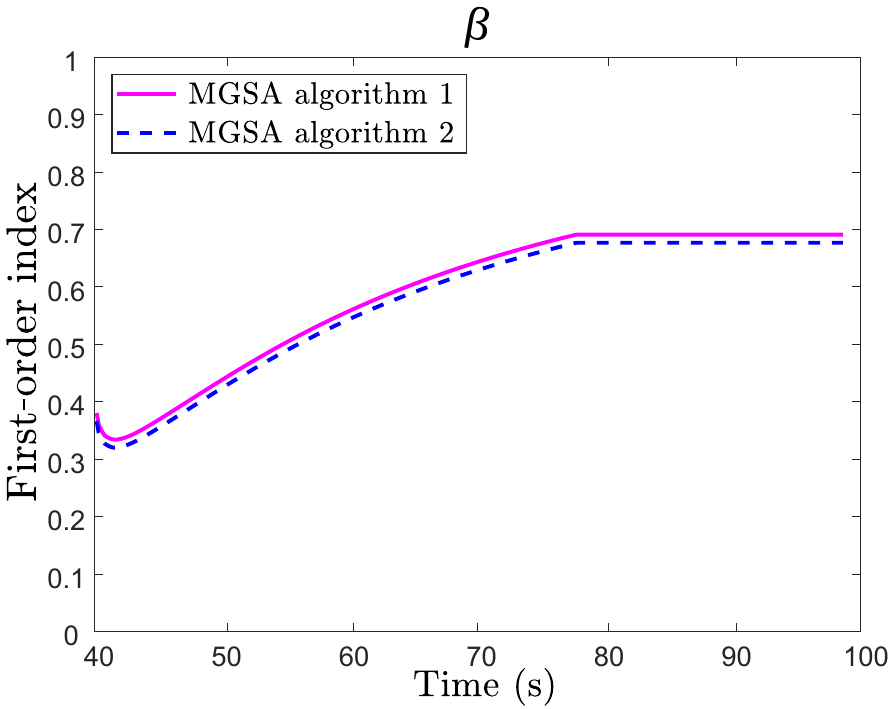}
      \caption{}
      \label{SensitivityIndices_BLModel20c}
    \end{subfigure}
    \begin{subfigure}{.45\textwidth}
      \includegraphics[width=\textwidth,keepaspectratio]{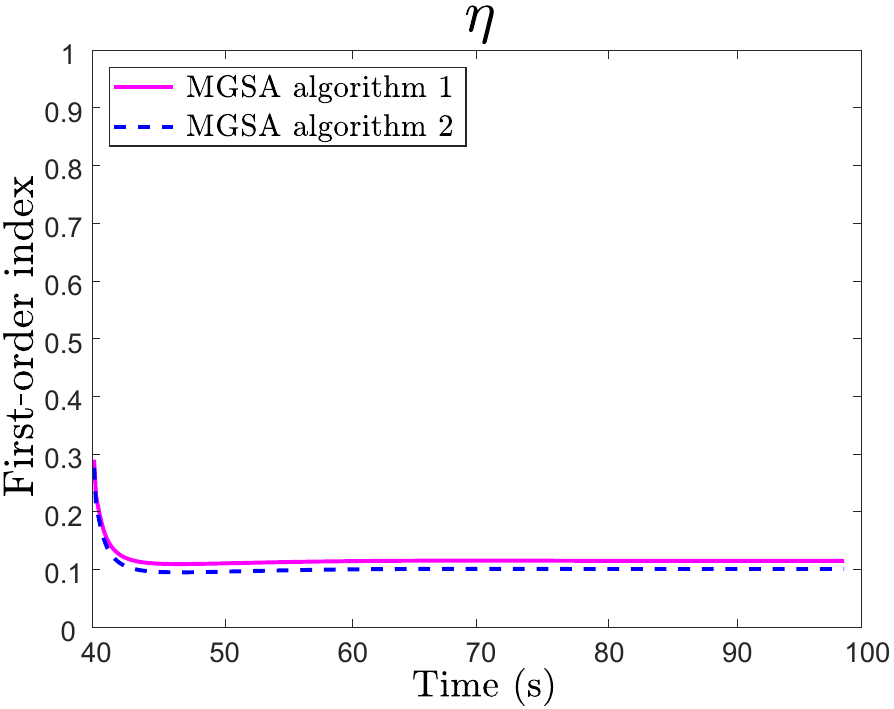}
      \caption{}
    \end{subfigure}
\end{center}
\caption{Sensitivity indices for the bond length between the 20th and 21st filaments in the second layer}
\label{SensitivityIndices_BLModel20}
\end{figure*}

\begin{figure}[!htb]
\centerline{\includegraphics[width=.4\columnwidth,keepaspectratio]{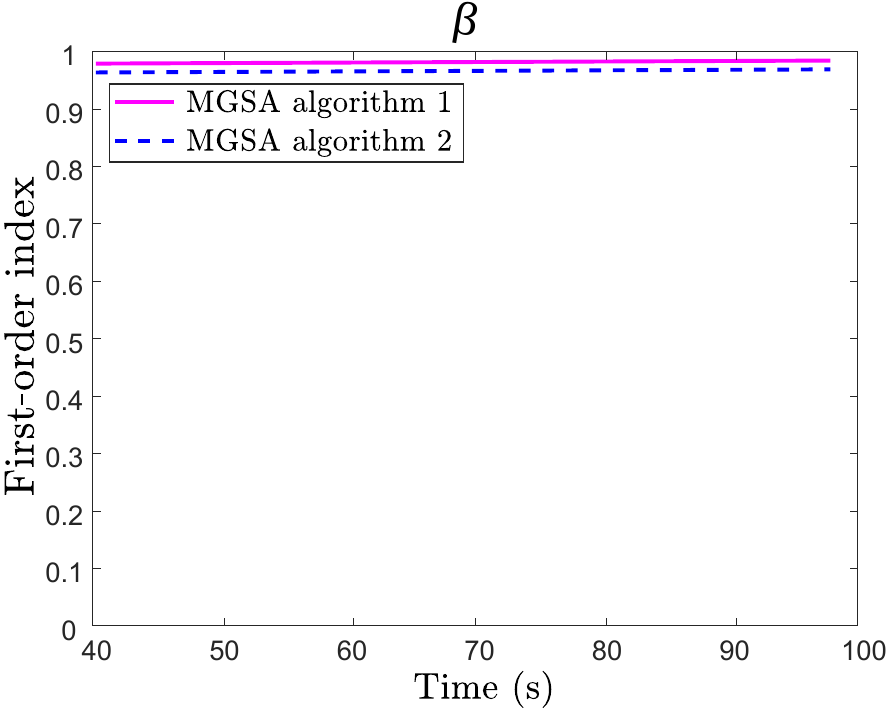}}
\caption{Sensitivity indices for the bond length between the 50th and 51st filaments in the fourth layer}
\label{SensitivityIndices_BLModel50}
\end{figure}

\begin{figure}[!htb]
\centerline{\includegraphics[width=.4\columnwidth,keepaspectratio]{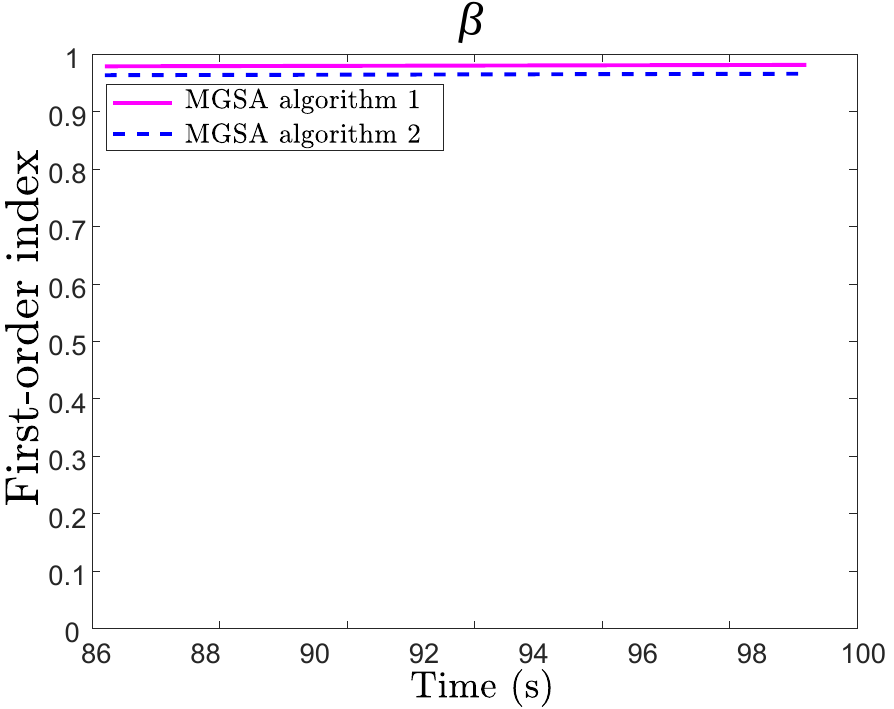}}
\caption{Sensitivity indices for the bond length between the 87th and 88th filaments in the sixth layer}
\label{SensitivityIndices_BLModel87}
\end{figure}

The coupled heat transfer and sintering neck growth model considers eleven parameters as uncertain, i.e., $\alpha$, $h_{conv}$, $\lambda_i$, $i\in\{1,2,3,4,5\}$, $\Gamma$, $\eta$, $\beta$ and $\gamma$. As discussed earlier, the contributions of various uncertainty sources to the neck growth vary for each layer as well; whereas, these contributions to the neck growth between filaments within a layer remain the same. Therefore, the first-order Sobol' indices of material parameters are assessed for four different neck growths at four different layers, i.e., the 1st, 20th, 50th and 87th bond formations in the first, second, fourth and sixth layers, respectively. We performed sensitivity analysis for each layer, at several interfaces within each layer. We present the results for only 4 layers (i.e., first, second, fourth and sixth layers) out of a total of 6 layers since the results corresponding to the fifth layer were the same as for the fourth and sixth layers, and the results corresponding to the third layer were the same as for the second layer. The results from GSA for these neck growths are illustrated in Figs.~\ref{SensitivityIndices_BLModel1},~\ref{SensitivityIndices_BLModel20},~\ref{SensitivityIndices_BLModel50}, and~\ref{SensitivityIndices_BLModel87}.

The contributions of $\beta$ increase significantly during the deposition of the specimen, while the contributions of $\alpha$, $\lambda_1$, $\lambda_2$, $\lambda_3$, $\lambda_4$, $\lambda_5$, $h_{conv}$, $\Gamma$, $\eta$ and $\gamma$ to the variations in the neck growth at given layers are negligible. The sensitivity index of $\beta$ increases as the temperature of the interface between the 1st and 2nd, and the 20th and 21st filaments cools down as illustrated in Figs.~\ref{SensitivityIndices_BLModel1c} and~\ref{SensitivityIndices_BLModel20c}. 

The uncertainty grows fast as the deposition time increases or temperature decreases because the influence of uncertainty in the model parameter $\beta$ on the neck growth increases with decreasing temperature. Another reason for the increase in uncertainty as the temperature decreases is that the neck growth model cannot capture the physics accurately when the temperature is below 130$^{\circ}$C as shown in Fig.~\ref{fig:CorrectedPred}. 

Based on the contributions of the various uncertainty sources to the neck growth at each layer of the part, $\beta$ was found to be dominant for all the layers and interfaces considered; next was $\alpha$ (considering the high contribution of $\alpha$ to the uncertainty in the temperature evolution of filaments) which was found to be significant in some of the layers. All the other parameters are fixed at their mean value to reduce computational effort since their contribution to the uncertainty in the neck growth predictions were negligible compared to the parameters $\alpha$ and $\beta$, based on the rigorous sensitivity analysis described in Section~\ref{Sec:GSA}.

\subsection{Model calibration}\label{Sec:Calibration}
In a Bayesian setting, the epistemic uncertainty regarding the model parameters that have significant sensitivity indices can be reduced using experimental data. As discussed in Section~\ref{Sec:GSA}, all material properties and model parameters, except $\alpha$ and $\beta$, are fixed at their nominal values. The measurement error is considered to be negligible since the measured bond lengths are precise to seven decimal points. Note that the material property $\alpha$, which is required for the heat transfer analysis, is not needed in case B, where experimentally measured temperature profile is the input to the sintering neck growth model instead of the heat transfer model prediction.

\begin{figure}[!htb]
\centerline{\includegraphics[width=.5\columnwidth,keepaspectratio]{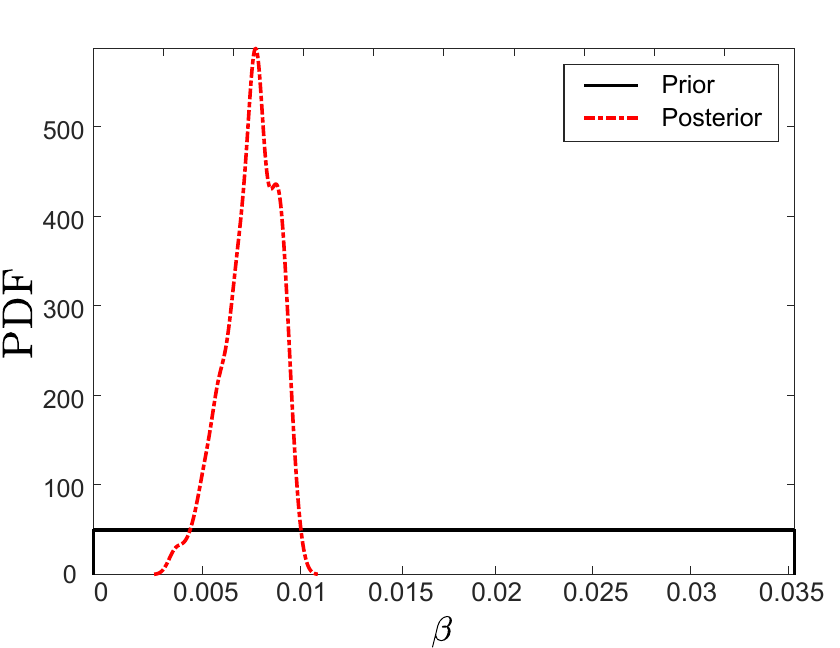}}
\caption{Prior and posterior distributions of the model parameter $\beta$ using experimental temperature profile as the input to the sintering neck growth model (case B)}
	\label{fig:MCMC_Exp_Beta}
\end{figure}

The posterior distribution for the sintering neck growth model parameter $\beta$ for case B (using observed temperature data as the input) is illustrated in Fig.~\ref{fig:MCMC_Exp_Beta}.
\begin{figure*}[!htb]
\begin{center}
\begin{subfigure}{.45\textwidth}
  \centering
  \includegraphics[width=\columnwidth,keepaspectratio]{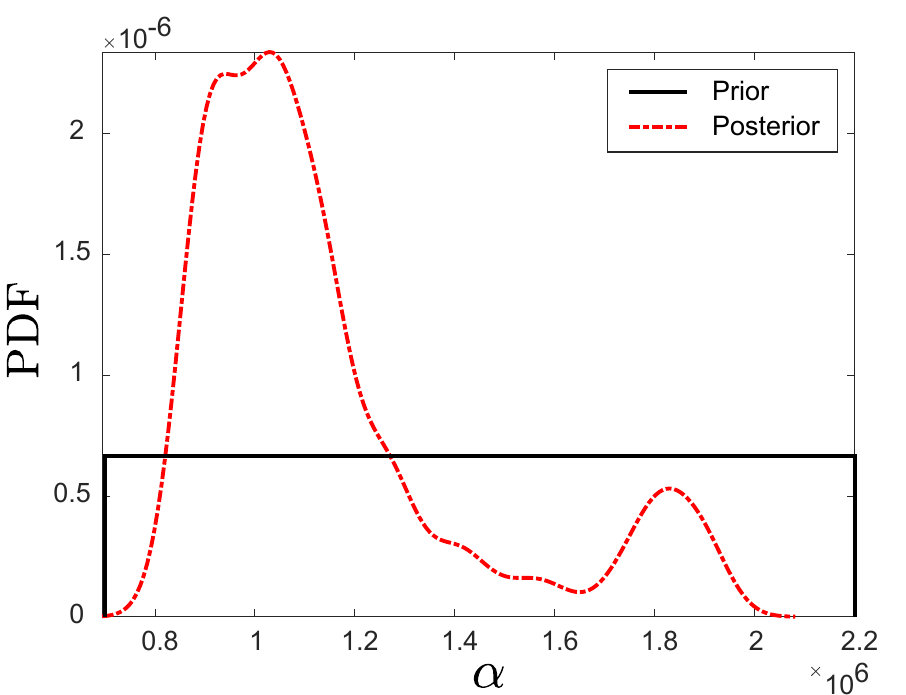}
  \caption{}
\end{subfigure}
\begin{subfigure}{.45\textwidth}
  \centering
  \includegraphics[width=\columnwidth,keepaspectratio]{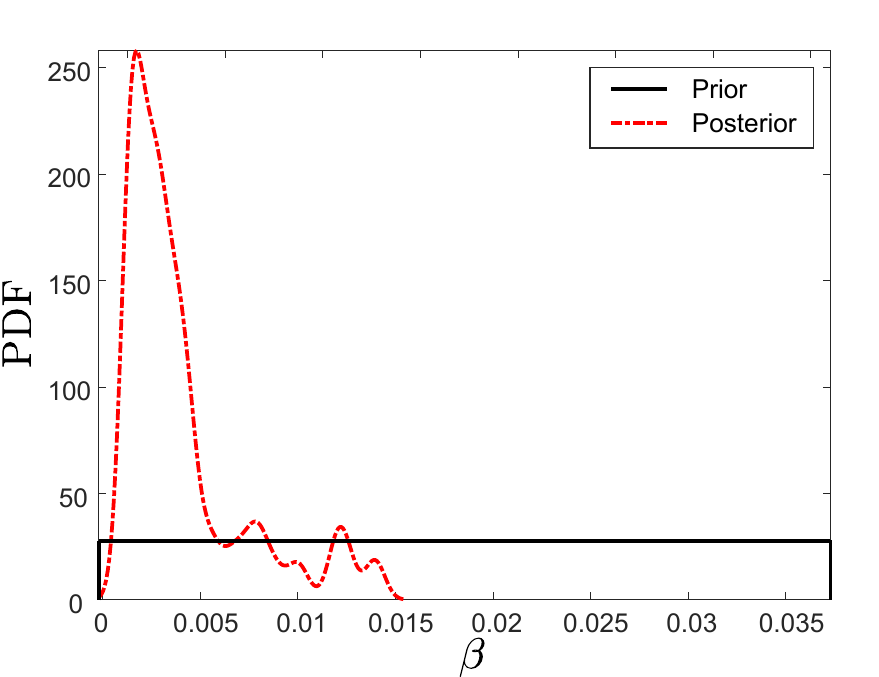}
  \caption{}
\end{subfigure}
\end{center}
\caption{Prior and posterior distributions of the material property $\alpha$ and model parameter $\beta$ considering the heat transfer model predictions as the input to the sintering neck growth model (case A)}
\label{fig:MCMC_Temp_alphabeta}
\end{figure*}
For case A, where heat transfer model predictions are the input to the sintering neck growth model, the posterior distributions for $\alpha$ and $\beta$ are shown in Fig.~\ref{fig:MCMC_Temp_alphabeta}.

In order to calibrate these model parameters, 15,000 posterior samples are drawn using MCMC and the initial 5,000 samples are rejected (initial burn-in samples). The last 10,000 samples yield a mean value of 1.196 $\times$ 10$^6$, and a coefficient of variation of 0.215 for $\alpha = \rho C$ and two different posterior distributions of $\beta$ with mean values of 0.00378 and 0.0193, and coefficient of variations of 0.1154 and 0.05 for case A and B respectively. The mean of the posterior distribution of $\beta$ for case B is close to the value reported in the literature~\cite{sun2008effect}. However, for case A the uncertainty in $\alpha$ has a significant effect on the posterior distribution of $\beta$ by shifting its mean to a smaller value. These values are used in the subsequent analysis (i.e., in the surrogate models for the model errors and optimization under uncertainty).

\subsection{Surrogate modeling}
The surrogate models for the model discrepancy in cases A and B are built using the calibrated model parameters illustrated in Section~\ref{Sec:Calibration}. The GP model described in Section~\ref{Sec:GP} is built with the training point inputs $[T_n, v_p, x, y]$ (i.e., nozzle temperature, printer speed, and the location of midpoint of the intra-layer bonds, respectively), and the corresponding intra-layer bond length model discrepancy $\boldsymbol{\delta}$. Then, for a given combination of nozzle temperature and speed, the predicted model discrepancy is used to correct the bond length estimated by the neck growth model.

\begin{figure}[!htb]
\centerline{\includegraphics[width=.6\columnwidth,keepaspectratio]{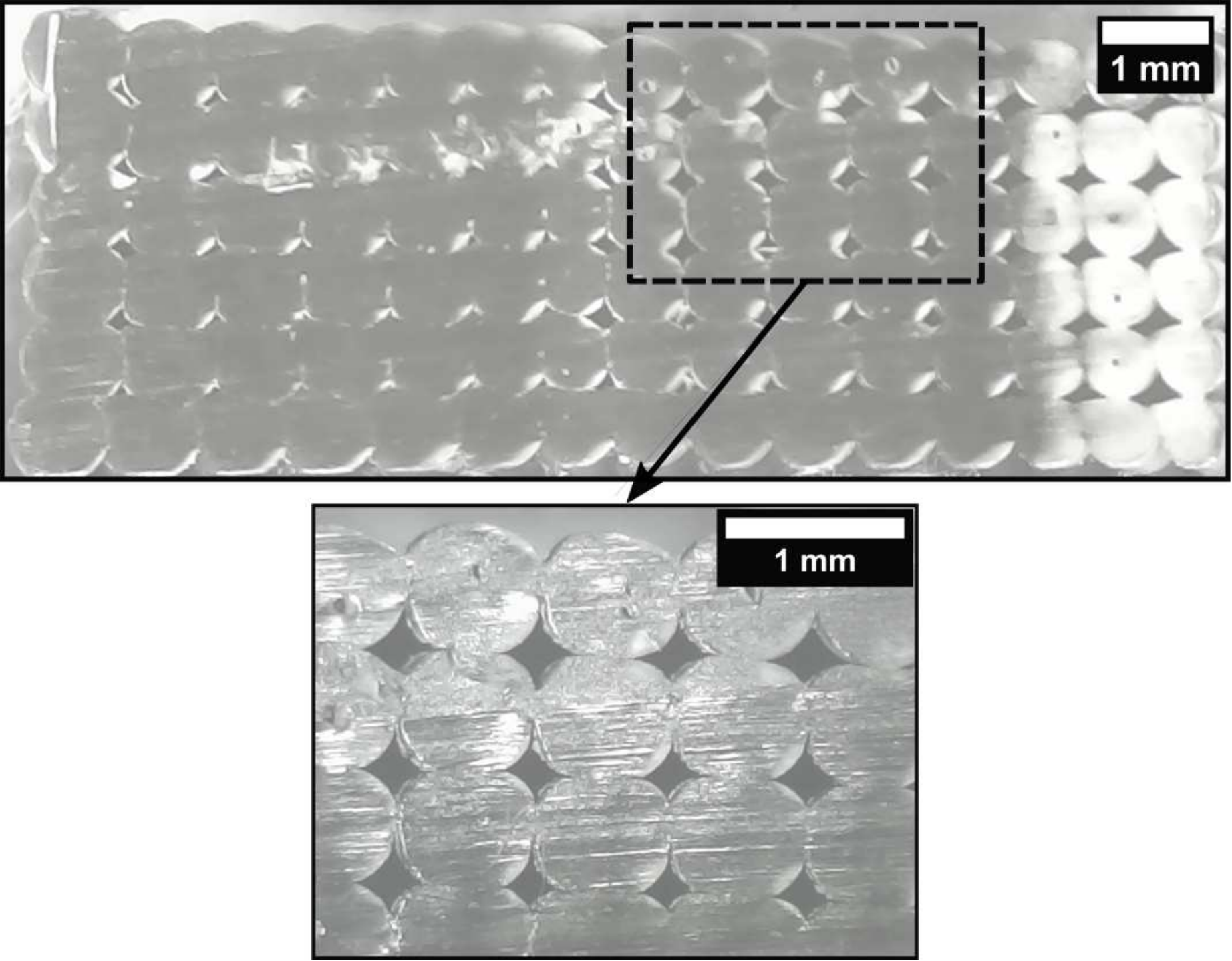}}
	\caption{Cross-section view of a sample produced with $T_n=$ 240$^{\circ}$C, and $v_p =$ 0.042\ m/s} 
	\label{fig:ExperimentalCuts}
\end{figure}

A series of experiments with different combinations of nozzle temperature and printer speed are used to measure the bond length, thus providing discrepancy data to train and test the above GP surrogate model. The specimens were sectioned at the midpoint, i.e. $z_{\rm cut}=L/2=$ 0.0175\ m, and their cross-sections were analyzed under a digital microscope. The features of these cross-sections were analyzed using the image processing program ImageJ~\cite{schneider2012nih} (Fig. \ref{fig:ExperimentalCuts}).

For a sample that is printed with inputs $(T_n,v_p)=$ (240$^{\circ}$C, 0.042\ m/s), the intra-layer bond lengths between adjacent filaments at $z_{\rm cut}=$ 0.0175 m of each layer are given in Table.~\ref{table:BL_240_42_1}. The numbering of the interfaces is done from left to right for all layers.\ For example, the label for the interface between 1st and 2nd filaments is 1 and the label for the interface between 89th and 90th filaments is 14.
\begin{table*}[!htb]
\tabcolsep=1pt\relax
\scriptsize
    \caption{Intra-layer bond length measurements at each layer for $(T_n,v_p)=($240$^{\circ}$C, 0.042\ m/s$)$}
    \label{table:BL_240_42_1}
\centering{%
    \begin{tabular*}{\linewidth}{@{\extracolsep{\fill}}l*{14}c@{\extracolsep{\fill}}}
    \toprule
            Layer &  1 & 2 & 3 & 4 & 5 & 6 & 7 & 8 & 9 & 10 & 11 & 12 & 13 & 14 \\
    \midrule
            1 & 0.4146 & 0.4558 & 0.5050 & 0.5050 & 0.5130 & 0.4339 & 0.4425 & 0.3915 & 0.3806 & 0.3930 & 0.3840 & 0.3798 & 0.3323 & 0.3162  \\
            2 & 0.4529 & 0.5056 & 0.5398 & 0.5357 & 0.5345 & 0.5050 & 0.3350 & 0.3293 & 0.3709 & 0.3833 & 0.4041 & 0.3854 & 0.2870 & 0.2606  \\
            3 & 0.4203 & 0.4828 & 0.5469 & 0.5517 & 0.5046 & 0.4889 & 0.3771 & 0.2269 & 0.2406 & 0.3271 & 0.3284 & 0.3361 & 0.2758 & 0.2506 \\
            4 & 0.4619 & 0.4895 & 0.5472 & 0.5944 & 0.5917 & 0.5196 & 0.3472 & 0.2055 & 0.2266 & 0.2608 & 0.3069 & 0.3316 & 0.2780 & 0.2781 \\
            5 & 0.4268 & 0.4798 & 0.5002 & 0.6125 & 0.5746 & 0.5667 & 0.4010 & 0.2617 & 0.1874 & 0.3071 & 0.2743 & 0.2913 & 0.2334 & 0.2621 \\
            6 & 0.4185 & 0.4694 & 0.4891 & 0.5334 & 0.4948 & 0.4750 & 0.3427 & 0.2617 & 0.2356 & 0.2886 & 0.2771 & 0.2670 & 0.2731 & 0.2884 \\
    \bottomrule
    \end{tabular*}
}%
\end{table*}

Using Latin hypercube sampling, 25 sets of process parameters were generated, and experiments were conducted at these 25 values. The experimental data is divided into 2 sets, namely training set (20 specimens) and testing set (5 specimens). The training set is further subdivided into two subsets for cross-validation; $k$-fold cross-validation is performed by splitting the training set into 16 sets for model training and 4 sets for cross-validation, and these sets are selected randomly $k=5$ different times. The bond length predictions using the observed temperature profiles are different than the ones using the heat transfer model predictions as inputs to the sintering neck growth model. Thus, two different GP models are built to represent the model error associated with these two cases, i.e., the bond length predictions using (A) heat transfer model predictions, and (B) observed temperature data. In each fold of training and cross-validation, the GP models are trained using 16 sets of data and the trained models are cross-validated using the remaining 4 sets of data. The average cross-validation accuracy of the GP models over the 5 folds (random shuffles) is assessed by evaluating the average mean squared error (MSE), which is found to be 0.0018 and 0.0017 for cases A and B respectively.

\begin{figure}[!htb]
\centerline{\includegraphics[width=.5\columnwidth,keepaspectratio]{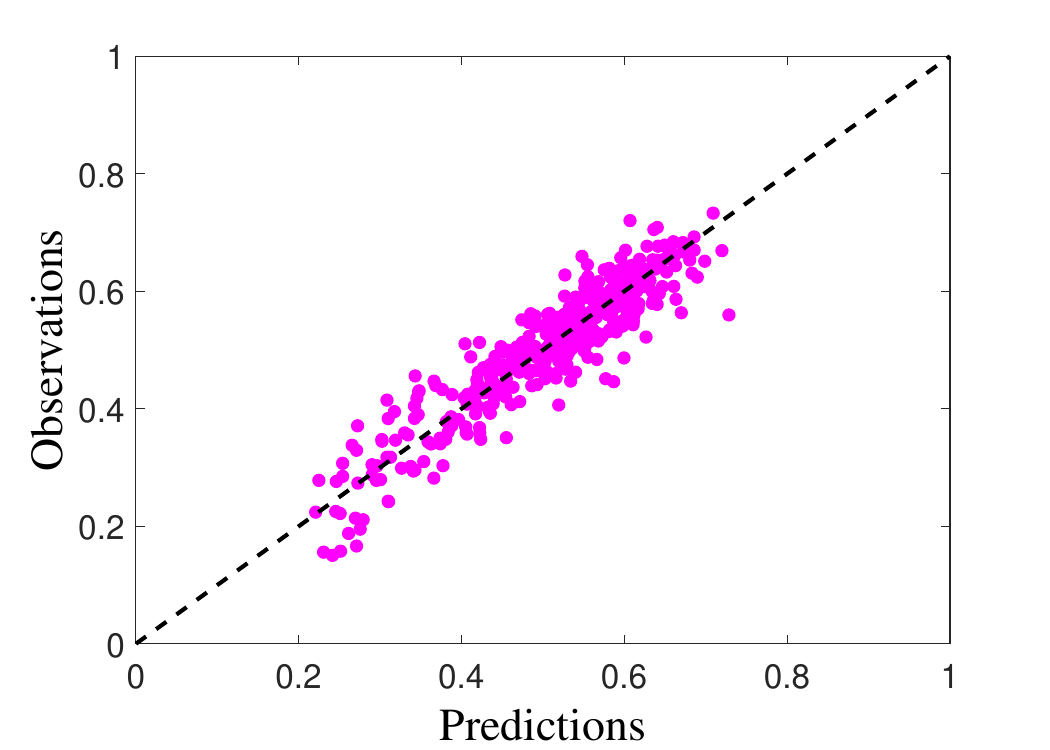}}
		\caption{Validation of the bond length prediction model for case A, where x and y-axes represent the bond length predictions and experimental observations respectively at each interface between adjacent filaments}
	\label{fig:pred_obs_a}
\end{figure}

\begin{figure}[!htb]
\centerline{\includegraphics[width=.5\columnwidth,keepaspectratio]{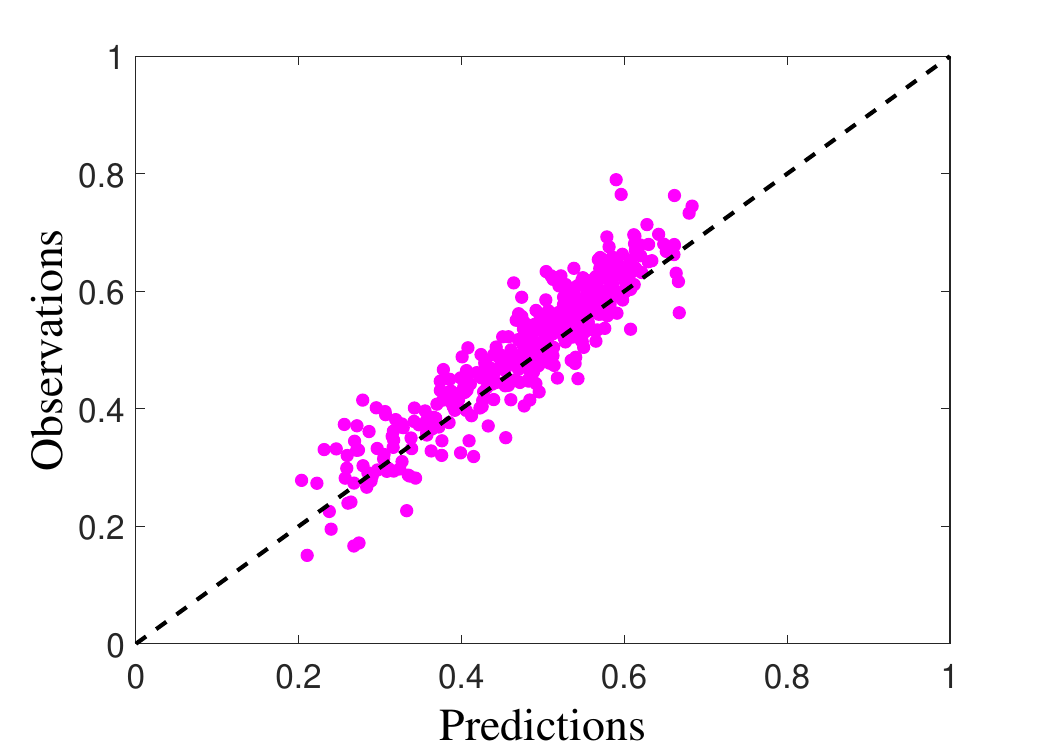}}
		\caption{Validation of the bond length prediction model for case B, where x and y-axes represent the bond length predictions and experimental observations respectively at each interface between adjacent filaments}
	\label{fig:pred_obs_b}
\end{figure}

Further validation of the corrected bond length prediction model is done using testing data, i.e., 5 sets of experiments. The prediction accuracy of the corrected model for the test set is assessed by evaluating the mean squared error (MSE), which is found to be 0.0020 and 0.0019 for cases A and B respectively. MSE of 1\% is used as the quantitative criterion for acceptance. The results indicate that the MSE values for both models are less than 1\%, thus they are accepted for further analysis. The corrected bond length predictions at each interface based on case A and B are validated by comparing with the bond length measurements at each interface between adjacent filaments of the test set (see Figs.~\ref{fig:pred_obs_a} and~\ref{fig:pred_obs_b}. Note that the points are close to the 45-degree line, showing good agreement between predictions and observations even for individual interfaces. 

\subsection{Process design optimization}
\begin{figure}[!htb]
\centerline{\includegraphics[width=.5\columnwidth,keepaspectratio]{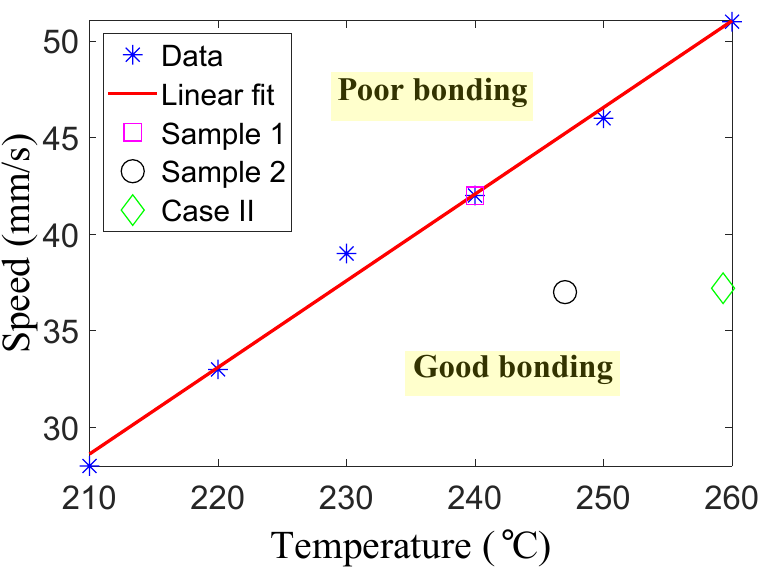}}
			\caption{Bonding frontier for Ultimaker 2 Extended +}
	\label{fig:BondFrontier}
\end{figure}

In this optimization, the process parameters (nozzle temperature and extrusion speed) are the design variables. The objective is to optimize the bond quality (indicated by bond length) at each layer of the specimen while satisfying the constraints on the design variables. The lower and upper bounds (LB and UB respectively) for the design variables are shown in Table~\ref{table:Bounds}. The lower and upper bounds have the same numerical values for all layers. The upper bound for the printer nozzle temperature was chosen as 260$^{\circ}$C, since the printer did not allow an extrusion temperature above 260$^{\circ}$C. The lower bound for the printer nozzle temperature was chosen as 210$^{\circ}$C because the quality of the specimens reduced significantly below a nozzle temperature of 210$^{\circ}$C. The lower and upper bounds for the printer extrusion speed $v_p$ were chosen as 0.015 and 0.043 m/s, respectively. We observed debonding and geometrical inaccuracies (warping) in the FFF parts that are printed using process parameter combinations above the linear fit. For example, a part printed with 220°C and 40 mm/s resulted in debonding in several layers and interfaces. These experimental data points and a linear fit to these combinations of process parameters that result in poor bonding (bonding frontier) are plotted in Fig.~\ref{fig:BondFrontier}. The overall bond quality of a part printed with parameter values above the bonding frontier is poor and delamination is observed. Whereas, the bond quality with parameter values below the bonding frontier is good, and gets better as the distance increases. 
\begin{table}[!htb]
\small
    \caption{Lower and upper bounds for the process design variables}
    \label{table:Bounds}
\centering{%
    \begin{tabular*}{\linewidth}{@{\extracolsep{\fill}}l*{2}c@{\extracolsep{\fill}}}
    \toprule
        Design variable & LB & UB \\
    \midrule
        $T_n$ (${^\circ}$C) & 210 & 260  \\
        $v_p$ (m/s) & 0.015 & 0.043 \\
    \bottomrule
    \end{tabular*}
}%
\end{table}

The methodology proposed in Section~\ref{Sec:Optim} is implemented here for the part shown in Fig.~\ref{fig:DeposSeqSequentiala} with $L=$ 0.035 m. The calibrated values of the material property $\alpha$ and model parameter $\beta$ in case A are used in the optimization of the neck growth at each layer. In order to demonstrate the robustness of the proposed formulation, two cases are considered: (I) different printer nozzle temperature and extrusion speed values for each layer, and (II) same printer nozzle temperature and extrusion speed value for all layers. The optimal solutions for these two cases are presented in Table~\ref{table:Design}.
\begin{table}[!htb]
\small
    \caption{Optimal printer nozzle temperatures~$({^\circ}$C) and extrusion velocities~(m/s)}
    \label{table:Design}
\centering{%
    \begin{tabular*}{\linewidth}{@{\extracolsep{\fill}}l*{4}c@{\extracolsep{\fill}}}
    \toprule
         & \multicolumn{2}{c}{Case I}  & \multicolumn{2}{c}{Case II} \\
        Layer & $T_n$ & $v_p$  & $T_n$ & $v_p$\\
    \midrule
        1 & 259.91 & 0.0430 & 259.30 & 0.0372 \\
        2 & 259.65 & 0.0162 & 259.30 & 0.0372 \\
        3 & 258.63 & 0.0150 & 259.30 & 0.0372 \\
        4 & 258.87 & 0.0150 & 259.30 & 0.0372 \\
        5 & 258.99 & 0.0150 & 259.30 & 0.0372 \\
        6 & 258.99 & 0.0150 & 259.30 & 0.0372 \\
    \bottomrule
    \end{tabular*}
}%
\end{table}

\begin{figure}[!htb]
\centerline{\includegraphics[width=.5\columnwidth,keepaspectratio]{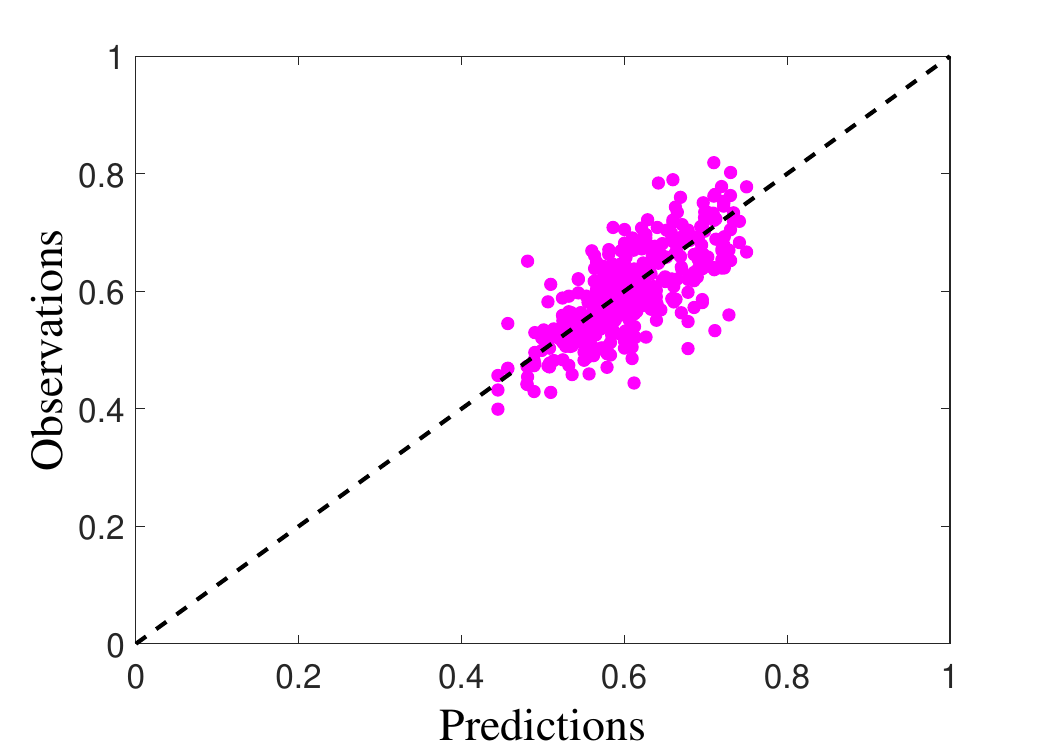}}
		\caption{Validation of the optimization results, where x and y-axes represent the bond length predictions and experimental observations respectively at each interface between adjacent filaments of Case I and II, respectively (three specimens printed for each case)}
	\label{fig:pred_obs_opt}
\end{figure}

\begin{figure}[!htb]
  \begin{center}
    \begin{subfigure}{.45\textwidth}
    \includegraphics[width=\textwidth,keepaspectratio]{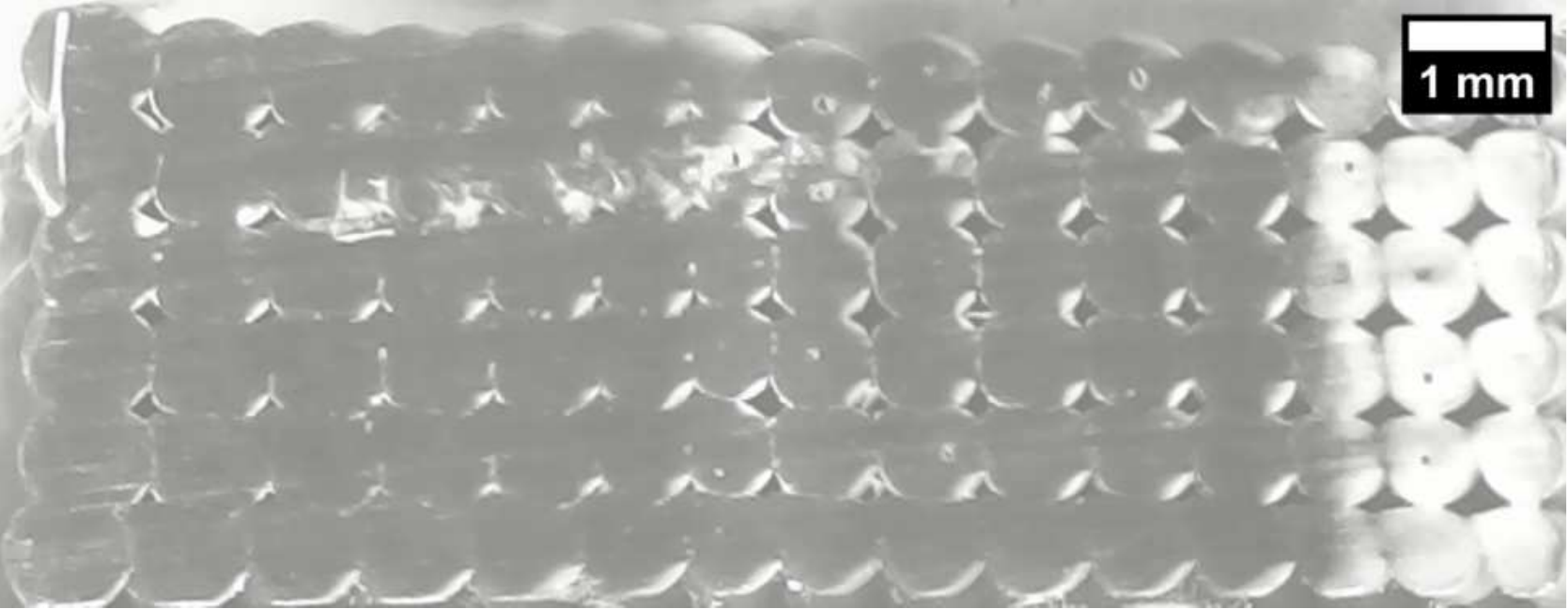}
      \caption{}
    \end{subfigure}
    \begin{subfigure}{.45\textwidth}
      \includegraphics[width=\textwidth,keepaspectratio]{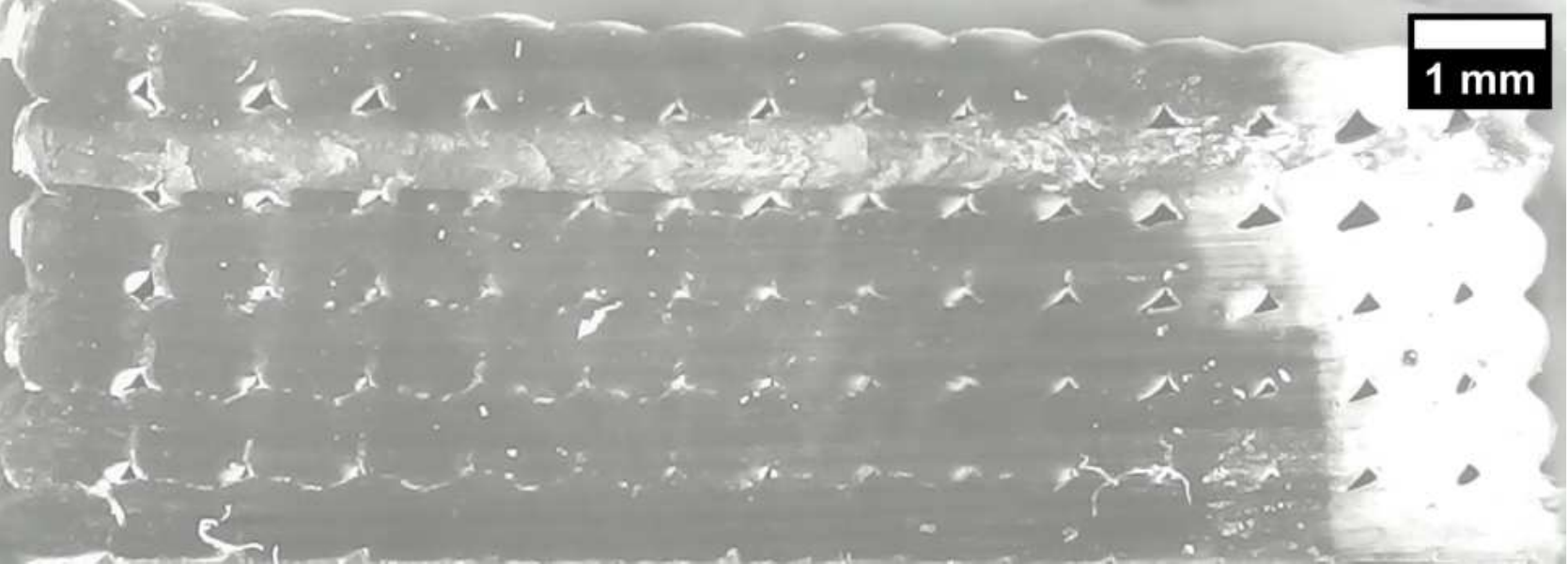}
      \caption{}
    \end{subfigure}
\\ 
    \begin{subfigure}{.45\textwidth}
    \includegraphics[width=\textwidth,keepaspectratio]{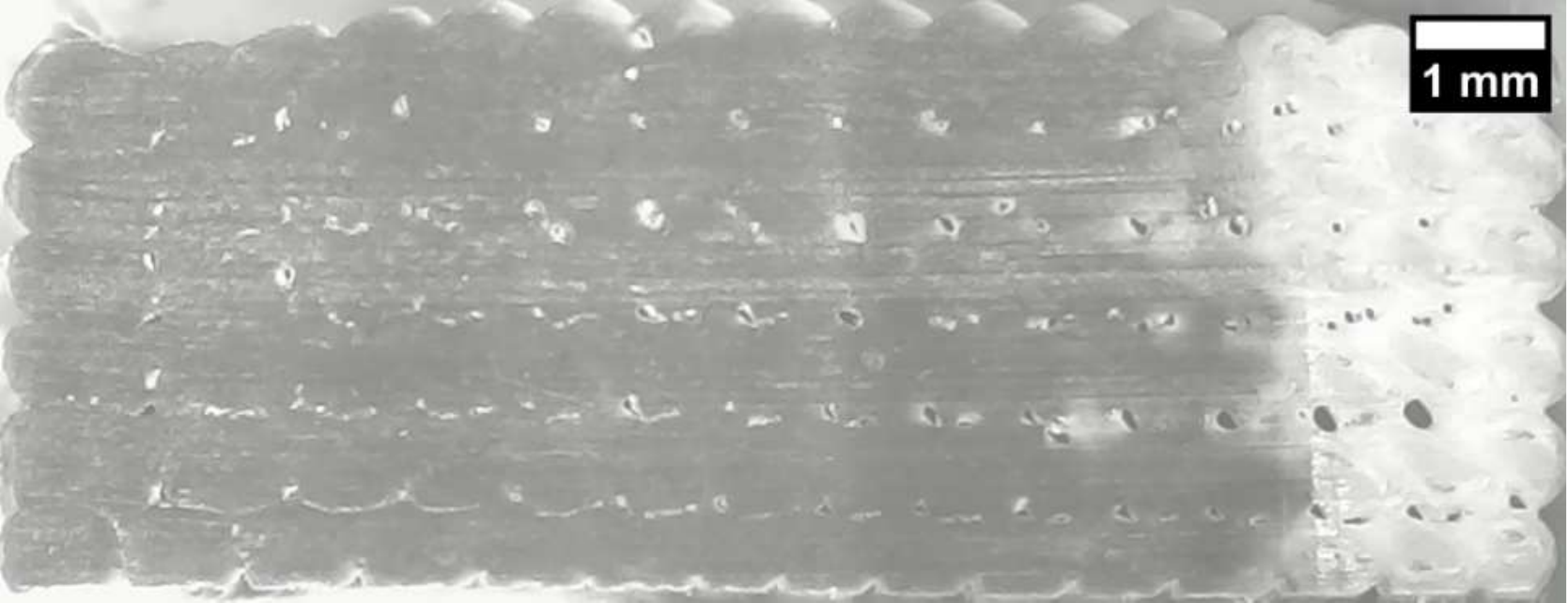}
      \caption{}
      \label{fig:PartsCrossSectionc}
    \end{subfigure}
    \begin{subfigure}{.45\textwidth}
    \includegraphics[width=\textwidth,keepaspectratio]{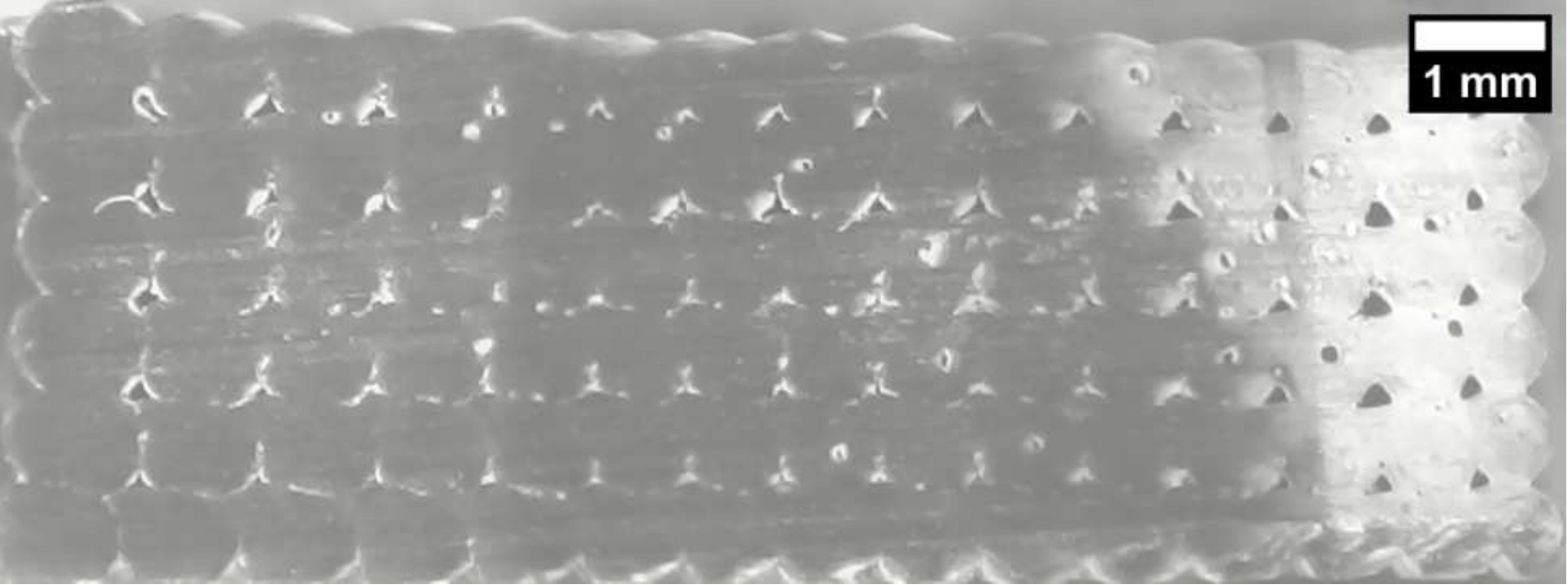}
      \caption{}
    \end{subfigure}
\end{center}
\caption{Cross-section views of the parts at $z_{\rm cut}=$ 0.0175 m built with non-optimal and optimal process parameters:~(a) sample 1,~(b) sample 2,~(c) case I and~(d) case II}
	\label{fig:PartsCrossSection}
\end{figure}

The optimal solutions are used to print three specimens for each case. The optimization results are then validated by comparing the bond length model predictions with the bond length measurements at each interface between adjacent filaments. The MSE value of 0.0027 is obtained based on the parts printed with the optimal solutions. Figure~\ref{fig:pred_obs_opt} shows the validation results for all six specimens, i.e., three specimens for each of Case I and II, printed with the corresponding optimum process parameter solution.
The mean and standard deviation of bond lengths of each layer are averaged across the three specimens ($\overline{\mu}_{\rm BL}$ and $\overline{\sigma}_{\rm BL}$). The standard deviation across the averaged mean of bond lengths of three parts $\sigma_{\rm parts}$ are shown in Table~\ref{table:BondQualOptimalSln} together with $\overline{\mu}_{\rm BL}$ and $\overline{\sigma}_{\rm BL}$. It is seen in Table~\ref{table:BondQualOptimalSln} that the variations across these specimens are relatively small. Note that the averaged mean of bond lengths ($\overline{\mu}_{\rm BL}$) across these three specimens is found to be smaller at the first layer than the other layers for the first case. The likely reasons for this difference are that the calibration/leveling of the build plate can be erroneous, and/or the printer extrusion speed decreases significantly (from 0.0430 m/s to 0.0162 m/s) as the second layer starts printing. These may result in excessive deformation on the first layer due to gravity and/or the weight of the material deposited above the first layer. As the bond quality starts reaching its upper limit (i.e., dimensionless neck radius $y/a=1$) as shown in Fig.~\ref{fig:PartsCrossSectionc}, the differences regarding the bond length between the top and bottom layers become negligible. Whereas, in the second case, since the dimensionless neck radius is still relatively less than unity, the decrease in the neck growth in the top few layers is more prominent. This difference can be attributed to the fact that the top layer cools down more than the bottom layers because a larger surface area of the filaments in the top layer is exposed to the environmental temperature. This results in relatively poor bonding in the top layers. As it can be seen in Table~\ref{table:BondQualOptimalSln}, the overall bond quality of the part and the bond quality at each layer are significantly better for the first case since the printer nozzle temperature and extrusion speed are optimized at each layer separately.
\begin{table}[!htb]
\small
    \caption{Overall average bond length at optimal solutions (all units are in millimeters)}
    \label{table:BondQualOptimalSln}
\centering{%
    \begin{tabular*}{\linewidth}{@{\extracolsep{\fill}}l*{6}c@{\extracolsep{\fill}}}
    \toprule
         & \multicolumn{3}{c}{Case I}  & \multicolumn{3}{c}{Case II} \\
        Layer & $\overline{\mu}_{\rm BL}$ & $\overline{\sigma}_{\rm BL}$ & $\sigma_{\rm parts}$ & $\overline{\mu}_{\rm BL}$ & $\overline{\sigma}_{\rm BL}$ & $\sigma_{\rm parts}$ \\
    \midrule
        1 & 0.56 & 0.0458 & 0.0479 & 0.58 & 0.0458 & 0.0049 \\
        2 & 0.65 & 0.0387 & 0.0292 & 0.59 & 0.0385 & 0.0053 \\
        3 & 0.65 & 0.0368 & 0.0146 & 0.59 & 0.0567 & 0.0022 \\
        4 & 0.64 & 0.0460 & 0.0086 & 0.57 & 0.0568 & 0.0149 \\
        5 & 0.66 & 0.0359 & 0.0068 & 0.55 & 0.0602 & 0.0129 \\
        6 & 0.65 & 0.0407 & 0.0048 & 0.52 & 0.0494 & 0.0081 \\
    \bottomrule
    \end{tabular*}
}%
\end{table}

Two specimens (sample 1 ($T_n=$ 240$^{\circ}$C and $v_p = $ 0.042\ m/s$)$ and sample 2 ($T_n=$ 247$^{\circ}$C and $v_p =$ 0.037\ m/s)) that are printed with non-optimal process parameters are used to demonstrate the effect of the proposed methodology. The total void area and the overall mean bond length (BL) at $z_{\rm cut}=$ 0.0175 of case I and II are compared with sample 1 and sample 2 in Table~\ref{table:Metric}. The cross-section views of these parts are shown in Fig.~\ref{fig:PartsCrossSection}. The overall bond quality of sample 2 represented by total void area (0.54 mm$^2$) and overall mean bond length metrics (0.52 mm) is better than the bond quality of sample 1 since the total void area and overall mean bond length of sample 1 are 2.11 mm$^2$ and 0.39 mm respectively (lower values of total void area and higher values of overall mean bond length imply a better-quality product). The total void area of the parts is identified with the use of microscopy images processed through the ImageJ software~\cite{schneider2012nih}, combined with a Matlab script to estimate the size of voids. The total void area is the smallest and the overall mean bond length is the largest for case I as expected, thus demonstrating the effectiveness of the proposed optimization methodology.
\begin{table}[!htb]
\small
    \caption{Total void areas and overall mean bond lengths at $z_{\rm cut}=$ 0.0175 m}
    \label{table:Metric}
\centering{%
    \begin{tabular*}{\linewidth}{@{\extracolsep{\fill}}l*{4}c@{\extracolsep{\fill}}}
    \toprule
        Metric & Case I & Case II & Sample 1 & Sample 2  \\
        \midrule
        Total void area ($\rm mm^2$) & 0.42 & 0.47 & 2.11 & 0.54  \\
        Overall mean BL ($\rm mm$) & 0.64 & 0.57 & 0.39 & 0.52 \\
    \bottomrule
    \end{tabular*}
}%
\end{table}

The coupled heat transfer and sintering neck growth model is directly used in the uncertainty quantification and optimization problems since the original simulation model is not very expensive ($\sim$100 s for one run on Intel\textsuperscript{\tiny\textregistered} Xeon\textsuperscript{\tiny\textregistered} CPU E5-2650 v4$@$2.20GHz with 64 GB RAM desktop machine).

\section{Conclusion}\label{Sec:Conclusion}
This paper developed a formulation for FFF process optimization under uncertainty, using an analytical solution for the transient heat transfer during filament deposition and cooling, and a sintering neck growth model. The neck growth between adjacent filaments is optimized while accounting for various sources of uncertainty and error. Variance-based sensitivity analysis is used to quantify the contribution of each uncertainty source to the variability of the output quantity (bond length). The physics model parameters that have the most significant contribution to the uncertainty in the model output are calibrated using experimental measurement of bond length. Additional experimental observations are used to build a surrogate model for the physics model discrepancy in predicting the bond length. The surrogate model is used to estimate the model discrepancy for given process parameter values and correct the physics model predictions. The corrected prediction model is used to select the optimal process parameters to maximize the bond quality at each layer. The optimum solution is validated with specimens printed with the optimized process parameters.

The proposed approach helps to replace the trial-and-error experimental approach with a model-based process parameter optimization under uncertainty in FFF. In future work, the proposed framework needs to be extended to online control of process parameters, thus further reducing the variability in the bond quality and other quantities of interest in order to achieve various product quality objectives. Current work did not include the effect of filament age, percentage of humidity in the filament, and ambient air flow and humidity; future work could include these effects. Future work can also explore the generalization capabilities of the proposed method to parts with more filaments, different geometries and sizes.

\section*{Acknowledgment} 
This study was supported by funds from the National Institute of Standards and Technology under the Smart Manufacturing Data Analytics Project (Cooperative Agreement No. 70NANB18H245). The support is gratefully acknowledged. At Vanderbilt University, valuable discussions with Dr. Pranav Karve and Dr. Paromita Nath, as well as support for experimental work by Garrett Thorne and Rich Tiesing are gratefully acknowledged.

\section*{Disclaimer}
Certain commercial equipment, instruments, or materials are identified in this paper in order to specify the experimental procedure adequately. Such identification is not intended to imply recommendation or endorsement by the National Institute of Standards and Technology, nor is it intended to imply that the materials or equipment identified are necessarily the best available for the purpose.

\bibliographystyle{unsrt}  
\bibliography{references}

\end{document}